\documentclass[11pt]{article} 
\usepackage[left=2.5cm, right=2.5cm, top=2.5cm, bottom=2.5cm]{geometry}

\usepackage{amssymb}
\usepackage{amsmath}

\usepackage[colorinlistoftodos]{todonotes}

\usepackage{listings} 
\usepackage{showexpl} 
\usepackage{color} 
\usepackage{amsthm} 
\usepackage{alltt}
\usepackage[export]{adjustbox} 

\usepackage[ruled,vlined]{algorithm2e}
\usepackage{algpseudocode}
\usepackage{amssymb}
\usepackage{amsmath}
\usepackage{amsfonts}
\usepackage{booktabs}
\usepackage{enumerate}
\usepackage{fancyhdr}
\usepackage{float}
\usepackage{graphicx}
\usepackage{lastpage}
\usepackage{mathrsfs}
\usepackage{authblk}
\usepackage{setspace}

\usepackage[colorlinks=true, citecolor=blue]{hyperref}
\hypersetup{
    colorlinks=true,
    linkcolor=blue,
    filecolor=magenta,      
    urlcolor=cyan,
}

\usepackage{appendix}
\usepackage{tikz}
\usetikzlibrary{decorations.text}
\usetikzlibrary{shapes,arrows}
\usepackage{pgfplots}
\usepackage{pgfplotstable}
\pgfplotsset{compat=newest}

\usetikzlibrary{calc}
\usetikzlibrary{spy}
\usetikzlibrary{positioning}
\usepackage{pdftexcmds}
\usetikzlibrary{external}
\usetikzlibrary{positioning}
\usetikzlibrary{arrows}
\usetikzlibrary{decorations.markings}
\usetikzlibrary{decorations.text}
\usetikzlibrary{backgrounds}
\usetikzlibrary{spy}
\usetikzlibrary{calc,patterns,decorations.pathmorphing,decorations.markings}
\usetikzlibrary{decorations.pathreplacing}
\usepgfplotslibrary{patchplots}

\usepackage{chngcntr}
\usepackage{etoolbox}
\usepackage{lipsum}

\usepackage{bm}

\makeatletter
\providecommand*{\input@path}{}
\g@addto@macro\input@path{{./}}
\makeatother

\makeatletter

\def\pgfplotstableread@openfile{%
    \def\pgfplotstable@loc@TMPa{\pgfutil@in@{ }}%
    \expandafter\pgfplotstable@loc@TMPa\expandafter{\pgfplotstableread@filename}%
    \ifpgfutil@in@
        \t@pgfplots@toka=\expandafter{\pgfplotstableread@filename}%
        \edef\pgfplotstableread@filename{\pgfplots@dquote\the\t@pgfplots@toka\pgfplots@dquote}%
    \fi
    \let\pgfplotstableread@old@crcr=\\%
    \def\\{\string\\}
    \openin\r@pgfplots@reada=\csname pgfk@/pgfplots/table file path\endcsname\pgfplotstableread@filename.tex
    \ifeof\r@pgfplots@reada
        \openin\r@pgfplots@reada=\csname pgfk@/pgfplots/table file path\endcsname\pgfplotstableread@filename\relax
    \else
        \pgfplots@warning{%
            You requested to open table '\pgfplotstableread@filename', but there is also a '\pgfplotstableread@filename.tex'. 
            TeX will automatically append the suffix '.tex', so I will now open '\pgfplotstableread@filename.tex'.
            Please make sure you don't accidentally load TeX files - this may produce unrecoverable errors.}%
        \closein\r@pgfplots@reada
        \openin\r@pgfplots@reada=\pgfplotstableread@filename\relax
    \fi
    \ifeof\r@pgfplots@reada
        \pgfplotsthrow{no such table file}{\pgfplots@loc@TMPa}{\pgfplotstableread@filename}{Could not read table file '\csname pgfk@/pgfplots/table file path\endcsname\pgfplotstableread@filename'. In case you intended to provide inline data: maybe TeX screwed up your end-of-lines? Try `row sep=crcr' and terminate your lines with `\string\\' (refer to the pgfplotstable manual for details)}\pgfeov%
        \global\let\pgfplotstable@colnames@glob=\pgfplots@loc@TMPa
        \def\pgfplotstableread@ready{0}%
    \fi
    \pgfplots@logfileopen{\pgfplotstableread@filename}%
    \let\\=\pgfplotstableread@old@crcr
}

\makeatother

\pgfplotscreateplotcyclelist{blue to red}{%
  color=blue\\%
  color=red!10!blue\\%
  color=red!20!blue\\%
  color=red!30!blue\\%
  color=red!40!blue\\%
  color=red!50!blue\\%
  color=red!60!blue\\%
  color=red!70!blue\\%
  color=red!80!blue\\%
  color=red!90!blue\\%
  color=red\\%
}
\pgfplotscreateplotcyclelist{red to blue}{%
  color=red\\%
  color=red!90!blue\\%
  color=red!80!blue\\%
  color=red!70!blue\\%
  color=red!60!blue\\%
  color=red!50!blue\\%
  color=red!40!blue\\%
  color=red!30!blue\\%
  color=red!20!blue\\%
  color=red!10!blue\\%
  color=blue\\%
}

\pgfplotscreateplotcyclelist{red to blue fast}{%
  color=red\\%
  color=red!80!blue\\%
  color=red!60!blue\\%
  color=red!40!blue\\%
  color=red!20!blue\\%
  color=blue\\%
}

\pgfplotscreateplotcyclelist{2red to blue}{%
  color=red\\%
  color=red\\%
  color=red!90!blue\\%
  color=red!90!blue\\%
  color=red!80!blue\\%
  color=red!80!blue\\%
  color=red!70!blue\\%
  color=red!70!blue\\%
  color=red!60!blue\\%
  color=red!60!blue\\%
  color=red!50!blue\\%
  color=red!50!blue\\%
  color=red!40!blue\\%
  color=red!40!blue\\%
  color=red!30!blue\\%
  color=red!30!blue\\%
  color=red!20!blue\\%
  color=red!20!blue\\%
  color=red!20!blue\\%
  color=red!10!blue\\%
  color=blue\\%
  color=blue\\%
}

\pgfplotsset{discard if/.style 2 args={x filter/.code={\ifnum\thisrow{#1}=#2\else\fi}}}

\usepackage{bm}

\usepackage{amsmath}

\usepackage{nicefrac}
\usepackage{multirow}
\usepackage{hhline}
\usepackage{lineno}
\usepackage{import}
\usepackage{epstopdf}
\usepackage{subcaption}
\usepackage{mathtools}
\usepackage{transparent}

\usepackage{listings}
\usepackage{cancel}
\usepackage{empheq}
\usepackage{parskip}

\usepackage{abstract}

\definecolor{tbf}{RGB}{255,0,0} 
\definecolor{txue}{RGB}{0,0,255}

\SetKwInput{KwInput}{Input}
\SetKwInput{KwOutput}{Output}

\newcommand{\ud}{\,\mathrm{d}}  

\usepackage{indentfirst}
\title{\Large Three-dimensional third medium contact model for hyperelastic contact and pneumatically actuated systems}

\begin{document}

\author[1]{Bing-Bing Xu}
\author[1]{\normalsize Tianju Xue
\footnote{\textit{cetxue@ust.hk} (corresponding author)}}
\affil[1]{\footnotesize Department of Civil and Environmental Engineering, The Hong Kong University of Science and Technology, Hong Kong, China}
\author[2]{Peter Wriggers}
\affil[2]{\footnotesize Institute of Continuum Mechanics, Leibniz University Hannover, Hannover, Germany}
\date{}
\maketitle
\vspace{-20pt}

\begin{abstract}
This work presents a comprehensive three-dimensional third-medium contact framework for modeling complex contact interactions in hyperelastic solids and pneumatically actuated systems.
The proposed third-medium formulation embeds a fictitious medium (or third medium) 
between potentially interacting bodies,
enabling a unified and robust treatment of hyperelastic contact 
and self-contact without the need for discretization of the contact interface.
Unlike the widely studied two-dimensional problem,
this paper extends the new regularization term given in Reference \cite{TMCWriggers2} to three-dimensional problems and ensures element quality in a third medium. 
Due to the need for higher-order elements for the regularization term, 
this paper details the linearization process of this problem within the finite element framework.
In addition, pneumatically actuated systems are considered by introducing
a pneumatic term to represent pneumatic loading (pressure or suction) and inducing contact caused by internal inflation.
This approach is suitable for complex hyperelastic contact and self-contact, 
and has potential applications in the fields of soft robotics and flexible mechanisms.
The framework is developed in a fully three-dimensional setting, 
making it also suitable for isogeometric methods and meshless methods.
Several benchmark and application-level simulations demonstrate the accuracy, robustness, and versatility of the proposed approach. 
The results highlight the capability of the three-dimensional third-medium model to handle challenging nonlinear contact scenarios relevant to soft materials, soft actuators, and emerging multifunctional structures.
\end{abstract}

\section{Introduction}
\label{p11.s1}
Contact phenomena involving large deformations arise in a wide range of engineering and scientific applications, 
particularly in soft materials, soft robotics, and pneumatically actuated systems \cite{Wriggers2006, Laursen2003}.
The accurately predicton of such interactions is essential for the design and optimization of advanced functional systems.
In classical computational contact mechanics,
various techniques for spatial discretization of contact have been developed, 
such as node-to-node, node-to-segment, and mortar methods \cite{Sauer2015,Mortar1,Mortar2}.
The resulting inequality constraints are typically enforced using methods like the Lagrange multiplier method, penalty method, 
augmented Lagrangian techniques, and Nitsche's formulation \cite{Wriggers2006,EDMcontact, Nitsche1}.
These methods rely on global and local contact searches and complex constraint enforcement strategies,
and have been successfully applied in both, small and large deformation contact problems \cite{Wriggers2006, Yastrebov2013}.
Additionally, self-contact problems have been addressed using similar approaches,
but special considerations are required to handle contact contributions on one of the contacting surfaces using projections from the other.

To avoid dealing with inequality constraints and searching for contact surfaces in contact mechanics,
a third medium contact approach was first proposed in \cite{TMCWriggers1} for two-dimensional large deformation contact problems.
In this approach, a third medium is introduced between the contacting bodies with an anisotropic constitutive behavior.
Then the contact problem is reformulated as a standard finite deformation problem,
where the third medium acts like a barrier to prevent penetration.
Since it eliminates the need to address contact interfaces, 
this technology can be used for many complex contact and self-contact problems.
The format is straightforward to implement within the finite element framework and has been extended to isogeometric analysis \cite{TMCcon3} and meshless methods \cite{TMCcon1}.
Besides, the technique has been successfully applied to three-dimensional engineering problems with higher-order finite elements for highly flexible cellular structures\cite{TMCcon2}. 
A similar technique named the contact domain method was proposed in \cite{CDM1, CDM2} for large deformation contact problems,
but still requires a global contact search due to the introduction of layer elements.

Later, the idea of the third medium contact method was extended to topology optimization problems involving contact.
As introduced in \cite{TMC3, TMC4, TMC5},
a regularization term is added to the strain energy density function of the third medium to stabilize highly distorted elements.
Different regularization terms have been proposed and discussed in \cite{TMC1, TMC5},
for example, the gradient of the deformation gradient \cite{TMC3, TMC4, TMC5} 
and the gradient of the skew-symmetric part of the deformation gradient \cite{TMCWriggers2}.
Besides, friction is taken into consideration in \cite{TMC6} using a crystal plasticity approach.
But as mentioned in \cite{TMC1, TMC5, TMC6}, the regularization term involves higher-order derivatives and requires higher-order elements,
which increases the computational cost.
Inspired by \cite{FirstOrder}, a new regularization technique was introduced in Wriggers et al. \cite{TMC2} 
by introducing auxillary fields for the computation of the high-order derivatives of the regularization term.
This leads to a formulation where first-order finite elements can be selected and two- and three-dimensional problems can be successfully handled \cite{TMCWriggers2}.

Another application of the third medium contact method is in pneumatically actuated systems.
Pneumatically actuated systems, such as soft robot actuators and fiber-reinforced actuators \cite{SoftRobots1,SoftRobots2}, 
experience pressure-driven inflation that naturally leads to geometric locking, folding, and self-contact phenomena.
Capturing these interactions in a unified and robust numerical model is essential for predictive simulations but remains an open research question.
Integration of pneumatic actuation within a third medium framework was first proposed in \cite{TMC1},
where a pneumatic term is added to the strain energy density function of the third medium.
However, this work has only considered two-dimensional problems.

In this work, we develop a general three-dimensional third-medium contact model tailored for hyperelastic contact and pneumatically actuated systems.
The regularization term is selected based on the skew-symmetric part of the deformation gradient proposed in \cite{TMCWriggers2}.
Different from the auxillary fields technique in \cite{TMCWriggers2},
the second-order ansatz space is employed directly to satisfy the requirement of higher-order derivatives in this work.
The linearization process of the third medium contact problem is detailed within the finite element framework.
Similar to \cite{TMC1} and \cite{TMC2}, a pneumatic term is incorporated to represent pressure or suction loading for 3D pneumatically actuated systems for the first time.

The paper is divided into the following parts.
In Section \ref{p11.s2}, we present the governing equations for the third medium contact model, 
including the strain energy density functions for the hyperelastic bodies and the regularization term for the third medium.
In Section \ref{p11.s3}, we derive the weak form and linearization of the third medium contact problem.
The finite element implementation is discussed in Section \ref{p11.s4}.
Several numerical examples are provided in Section \ref{p11.s5} to demonstrate the accuracy and robustness of the proposed approach.
Finally, conclusions are drawn in Section \ref{p11.s6}.


\section{Governing equations for contact mechanics using a third medium}
\label{p11.s2}
\subsection{Continuum model}
\label{p11.s2.1}
\begin{figure}[htbp]
\centering
\includegraphics[width=1.0\textwidth]{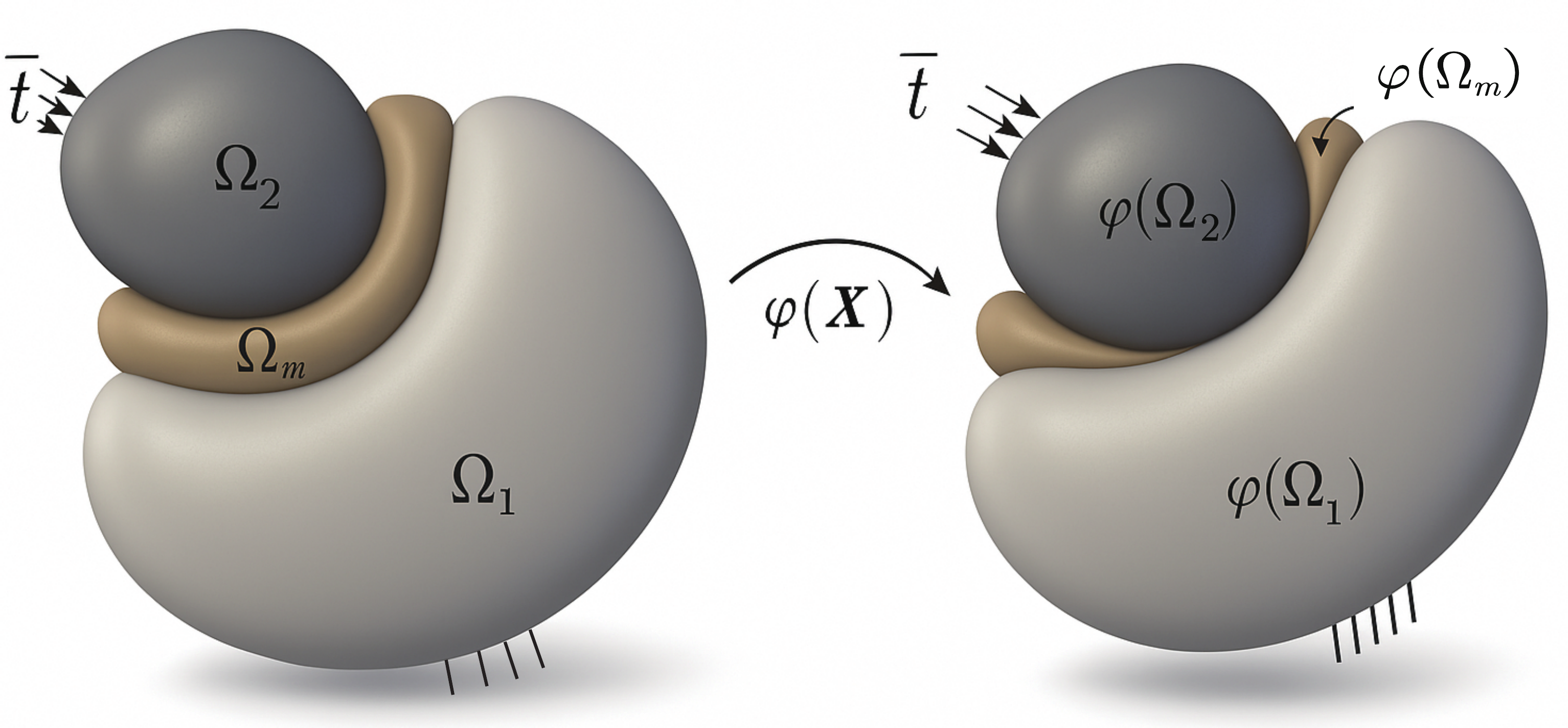}
\caption{Contact of two bodies with a third medium, where $\varphi$ is a map between the initial and current configuration.}
\label{p11.s2.tmc}
\end{figure}

Under the finite deformation assumption, we define a map $\varphi$ between the initial configuration $\Omega$ and the current configuration $\varphi(\Omega)$.
The deformation gradient is given by $\bm{F}=\nabla\varphi=\frac{\partial \varphi}{\partial \bm{X}}$,
where $\bm{X}$ is the coordinate in the initial configuration as shown in Fig.\ref{p11.s2.tmc}.
Then the right Cauchy–Green deformation tensor can be computed as $\bm{C}=\bm{F}^T\bm{F}$.
Using the split in volumetric and deviatoric terms, the strain energy density function for hyperelastic materials can be expressed as 
\begin{equation}
    \label{p11.s2.strainEnergy}
    \Psi(\bm{C}) = \frac{K}{2}\left(\ln J\right)^2+\frac{\mu}{2}\left(J^{-\frac{2}{3}}\text{tr}(\bm{C})-3\right),
\end{equation}
where $K$ is the bulk modulus, and $\mu$ is the shear modulus, $J=\det(\bm{F})$ is the volume change, see e.g. \cite{Wriggers2008}.
The form of the strain energy density function can be used to describe the contact bodies $\Omega_1$, $\Omega_2$ and the third medium $\Omega_m$,
however with different constitutive parameters.

The motion of the solid is governed by the principle of a stationary elastic potential
\begin{equation}
    \label{p11.s2.potential}
    \Pi = \int_\Omega\left[\Psi(\bm{C})-\bar{\bm{b}}\cdot\bm{u}\right]\ud\Omega-\int_{\Gamma_N}\bm{\bar{t}}\cdot\bm{u}\ud\Gamma_N\Rightarrow STAT,
\end{equation}
where $\bar{\bm{b}}$ and $\bar{\bm{t}}$ are the body forces and surface tractions, respectively.

Different from the conventional contact algorithms,
the basic idea of the third medium contact (TMC) is to introduce a third medium $\Omega_m$ as a layer between two bodies $\Omega_1$ and $\Omega_2$, as shown in Fig.\ref{p11.s2.tmc}.
It is clear that the stiffness of the third medium should be very small before contact and then rapidly increase once the two bodies are close to each other to prevent penetration.
To achieve the above objectives, we need to modify the strain energy density of the third medium.

\subsection{Strain energy density for the third medium}
\label{p11.s2.2}
As discussed before, the third medium should have almost no inﬂuence on the deformation of the two bodies before contact,
and becomes very stiff when the two bodies are close to each other.
In order to meet the above requirements, 
the strain energy density function $\Psi_m(\bm{u})$ for the third medium can be selected as, see e.g. \cite{TMC3}
\begin{equation}
    \label{p11.s2.Psim}
    \Psi_m(\bm{C}) = \gamma\left[\frac{K}{2}\left(\ln J\right)^2+\frac{\mu}{2}\left(J^{-\frac{2}{3}}\text{tr}(\bm{C})-3\right)\right].
\end{equation}
By selecting a small parameter $\gamma$, the third medium can act as a highly soft material before contact.
Since the contact between the solids is equivalent to the third medium being compressed to zero volume,
the term $\ln J$ has the property $J\rightarrow 0$ and $\Psi_m\rightarrow\infty$.
Thus, the third medium acts as a barrier and can avoid penetration when the bodies approach each other.

At the same times, the elements in the third medium $\Omega_m$ are highly distorted when the two bodies come close to each other,
especially when the contact problem has free boundaries.
As discussed in \cite{TMC1}, a regularization term $\Psi_r$ is introduced to control the element distortions automatically.
\begin{equation}
    \Psi^{TM} = \Psi_m+\Psi_r.
\end{equation}
There are several options for choosing regularization terms.
As introduced in \cite{TMC5}, the regularization term can be expressed as
\begin{equation}
    \label{p11.s2.3.Psir}
    \Psi_r =\alpha_r\frac{\gamma}{2}\left(\nabla \bm{F}\vdots\nabla\bm{F}-\frac{1}{n_{dim}}\text{Div}(\nabla\bm{u})\cdot\text{Div}(\nabla\bm{u})\right)
\end{equation}
where $n_{dim}$ is the dimension of the problem,
$\alpha_r$ is another parameter which controls the effect of the regularization term.
Besides, $\nabla\bm{F}$ is the gradient of the deformation gradient
\begin{equation}
    \label{p11.s2.3.nablaF}
	\nabla\bm{F}\vdots \nabla\bm{F} = \frac{\partial^2 u_i}{\partial X_j\partial X_k}\frac{\partial^2 u_i}{\partial X_j\partial X_k}.
\end{equation}
Here, repeated subscripts represent summation ($i,j,k=1\sim n_{dim}$).
It should be noted that Eq.\eqref{p11.s2.3.nablaF} requires calculating the second derivative of the displacement with respect to the coordinates, 
therefore at least a second-order FEM is needed.

Generally, the regularization term is dominated by the shearing portion of the deformation gradient.
For three-dimensional problems, another regularization term is suggested by Wriggers \cite{TMCWriggers2} as 
\begin{equation}
    \label{p11.s2.3.fskew}
    \Psi_r = \alpha_r\frac{\gamma}{2}\nabla\bm{f}^{skew}\vdots \nabla\bm{f}^{skew}
\end{equation}
where 
\begin{equation}
    \bm{f}^{skew} = \frac{1}{2}\left(\bm{F}-\bm{F}^T\right).
\end{equation}
Compared with the regularization term in Eq.\eqref{p11.s2.3.Psir}, 
the regularization term is simpler (three components define the skew symmetric part) 
and needs less evaluation time than the regularization term Eq.\eqref{p11.s2.3.Psir}. 
In this work, we will use the newly proposed regularization term in Eq.\eqref{p11.s2.3.fskew} for simulation.

\subsection{Pneumatic term}
For the pneumatically actuated systems, another energy term $\Psi_p$ should be added to the strain energy density function,
so that the hydrostatic Cauchy stress across the third medium material is exactly equal to a prescribed pressure value \cite{Ogden2013}:
\begin{equation}
    \label{p11.s2.Psitmc2}
    \Psi^{TM}(\bm{u}) =\Psi_m(\bm{u})+\Psi_r(\bm{u}) +\Psi_p.
\end{equation}
The pneumatic part $\Psi_p$ can be selected as 
\begin{equation}
    \label{p11.s2.2.PsiP}
    \Psi_p = \bar{P}J,
\end{equation}
where $\bar{P}$ is the pressure/suction.
As illustrated in \cite{TMC1}, the Cauchy stress can be obtained as 
\begin{equation}
    \bm{\sigma} = \frac{1}{J}\bm{P}\cdot\bm{F}^T 
    = \frac{1}{J}\frac{\partial \Psi_p}{\partial \bm{F}}\cdot\bm{F}^T = \bar{P}\bm{I},
\end{equation}
where $\bm{I}$ is the second order identity tensor.
Then, suction corresponds to a positive pressure $\bar{P}$, while inflation is a result of a negative pressure $\bar{P}$.

By introducing this strain energy density, hydrostatic pressure can be generated in the third medium.
With regularization terms we can automatically control the element distortions during the simulation of pneumatically actuated systems and contact problems.
Next, we will present the finite element discretization of the third medium approach.

\section{Weak form and linearization}
\label{p11.s3}
Based on the above analysis, we find that the classical contact problem with inequality constraints has been reduced to a single finite deformation problem.
For the hyperelastic bodies $\Omega_1$ and $\Omega_2$ (denoted as $\Omega_s$), 
their strain energy densities are functionals of $\bm{C}$ or $\bm{F}$, see Eq.\eqref{p11.s2.strainEnergy}. 
However, for the third medium, due to the need to introduce regularization and pneumatic terms, 
its strain energy density is a functional of $\bm{F}$ and $\nabla\bm{F}$, see Eq.\eqref{p11.s2.3.fskew}. 
Therefore, it needs to be analyzed separately.

\subsection{Weak form and linearization for the hyperelastic body}
\label{p11.s3.1}
For the hyperelastic bodies $\Omega_s$,
the potential energy has the form as
\begin{equation}
    W = \int_{\Omega_s}\Psi(\bm{F})\ud \Omega
\end{equation}
The variational of the strain energy yields
\begin{equation}
    \label{p11.s3.1.deltaW}
    \delta W = \int_{\Omega_s}\frac{\partial\Psi(\bm{F})}{\partial\bm{F}}:\delta\bm{F}\ud \Omega = \int_{\Omega_s}\bm{P}:\delta\bm{F}\ud \Omega,
\end{equation}
where $\bm{P}$ is the first Piola-Kirchhoff stress tensor.
The linearization of Eq.\eqref{p11.s3.1.deltaW} can be expressed as
\begin{equation}
    \label{p11.s3.1.W2}
    \Delta(\delta W) = \int_{\Omega_s}\delta\bm{F}:\frac{\partial^2\Psi(\bm{F})}{\partial\bm{F}\partial\bm{F}}:\Delta\bm{F}\ud \Omega =
    \int_{\Omega_s}\left[\delta\bm{F}:\mathbb{C}^{PK1}:\Delta\bm{F}\right]\ud \Omega,
\end{equation}
where $\mathbb{C}^{PK1}$ is the fourth-order material tangent moduli tensor.
For the given strain energy density function in Eq.\eqref{p11.s2.strainEnergy},
the formualtions of the first Piola-Kirchhoff stress tensor $\bm{P}$ and the material tangent moduli tensor $\mathbb{C}^{PK1}$ 
can be obtained based on the techniques in \cite{Wriggers2008}.
In addition, automatic differentiation (AD) is also a convenient tool, see \cite{Korelc2016}, for the derivation of $\bm{P}$ and $\mathbb{C}^{PK1}$.

For ease of matrix representation, we can rewrite Eq.\eqref{p11.s3.1.deltaW} as
\begin{equation}
	\delta W = \int_{\Omega_s}\frac{\partial \Psi}{\partial\bm{E}}: \delta\bm{E}\ud\Omega 
    = \int_{\Omega_s}\bm{S}:\delta\bm{E}\ud\Omega,
\end{equation}
where $\bm{E}$ is the Green-Lagrange strain tensor, 
and $\bm{S}$ is the second Piola-Kirchhoff stress tensor with
\begin{equation}
    \bm{E} = \frac{1}{2}\left(\bm{F}^T\cdot\bm{F}-\bm{I}\right),
\end{equation}
where $\bm{I}$ is the second order identity tensor.
Then, the linearization can be expressed as
\begin{equation}
	\begin{aligned}
        \Delta(\delta W) &= \int_{\Omega_s}\delta\bm{E}:\frac{\partial^2\Psi}{\partial\bm{E}\partial\bm{E}}:\Delta\bm{E}\ud\Omega+\int_{\Omega_s}\Delta(\delta\bm{E}):\bm{S}\ud\Omega\\
        & = \int_{\Omega_s}\delta\bm{E}:\mathbb{C}^{PK2}:\Delta\bm{E}\ud\Omega+\int_{\Omega_s}\Delta(\delta\bm{E}):\bm{S}\ud\Omega.
    \end{aligned}
\end{equation}
where $\mathbb{C}^{PK2}$ describes the deformation depended material stiffness.
The formualtions of the second Piola-Kirchhoff stress tensor $\bm{S}$ and the material tangent moduli tensor $\mathbb{C}^{PK2}$ 
can also be obtained based on the techniques in \cite{Wriggers2008} or AD \cite{Korelc2016}.

\subsection{Weak form and linearization for the third medium}
\label{p11.s3.2}
The potential energy for the third medium $\Omega_m$ in Eq.\eqref{p11.s2.Psitmc2} is a functional of $\bm{F}$ and $\nabla \bm{F}$
\begin{equation}
    \label{p11.s3.2.W}
	W^{TM}(\bm{F},\nabla\bm{F}) = \int_{\Omega_m}\Psi^{TM}(\bm{F},\nabla\bm{F})\ud\Omega.
\end{equation}
Its variation can be expressed as
\begin{equation}
    \label{p11.s3.2.W1}
	\begin{aligned}
        \delta W^{TM} = \int_{\Omega_m}
	\left[\frac{\partial \Psi^{TM}}{\partial\bm{F}}:\delta\bm{F}+\frac{\partial \Psi^{TM}}{\partial\nabla\bm{F}}\vdots\delta\nabla\bm{F}\right]
	\ud\Omega = \int_{\Omega_m}
	\left(\bm{P}:\delta\bm{F}+\bm{T}\vdots\delta\nabla\bm{F}\right)
	\ud\Omega,
    \end{aligned}
\end{equation}
where 
\begin{equation}
    \bm{P} = \frac{\partial \Psi^{TM}}{\partial\bm{F}},\quad 
    \bm{T} = \frac{\partial \Psi^{TM}}{\partial\nabla\bm{F}}.
\end{equation}

For linearization, we consider the second variation of the potential energy (Eq.\eqref{p11.s3.2.W})
\begin{equation}
    \label{p11.s3.2.W2}
	\begin{aligned}
		\delta^2 W^{TM} &=\int_{\Omega_m}
		\delta\bm{F}:\frac{\partial^2\Psi^{TM}}{\partial\bm{F}\partial\bm{F}}:\delta\bm{F}+
		\delta\nabla\bm{F}\vdots\frac{\partial^2\Psi^{TM}}{\partial\nabla\bm{F}\partial\bm{F}}:\delta\bm{F}\\
		&
		+\delta\bm{F}:\frac{\partial^2\Psi^{TM}}{\partial\bm{F}\partial\nabla\bm{F}}\vdots\delta\nabla\bm{F}
		+\delta\nabla \bm{F}\vdots\frac{\partial^2\Psi^{TM}}{\partial\nabla\bm{F}\partial\nabla\bm{F}}\vdots\delta\nabla\bm{F}
		\ud\Omega\\
        &=\int_{\Omega_m}\left(
            \delta\bm{F}:\mathbb{C}:\delta\bm{F}
        +\delta\nabla\bm{F}\vdots\mathbb{A}:\delta\bm{F}
        +\delta\bm{F}:\mathbb{A}\vdots\delta\nabla\bm{F}
        +\delta\nabla \bm{F}\vdots\mathbb{B}\vdots\delta\nabla\bm{F}
        \right)\ud\Omega,
	\end{aligned}
\end{equation}
where 
\begin{equation}
	\mathbb{C} = \frac{\partial^2\Psi^{TM}}{\partial\bm{F}\partial\bm{F}},\quad
	\mathbb{A} = \frac{\partial^2\Psi^{TM}}{\partial\nabla\bm{F}\partial\bm{F}},\quad
	\mathbb{B} = \frac{\partial^2\Psi^{TM}}{\partial\nabla\bm{F}\partial\nabla\bm{F}}.
\end{equation}
Here, $\mathbb{C}$ is a fourth-order tensor, $\mathbb{A}$ is a fifth-order tensor, and $\mathbb{B}$ is an sixth-order tensor.
Again, automatic differentiation (AD) can be used to obtain these high-order tensors \cite{Korelc2016}.
If we disregard regularization terms and pneumatic terms, Eq.\eqref{p11.s3.2.W2} will degenerate into Eq.\eqref{p11.s3.1.W2}.

\section{Finite element implementation}
\label{p11.s4}
\begin{figure}[htbp]
    \centering
    \includegraphics[width=0.8\textwidth]{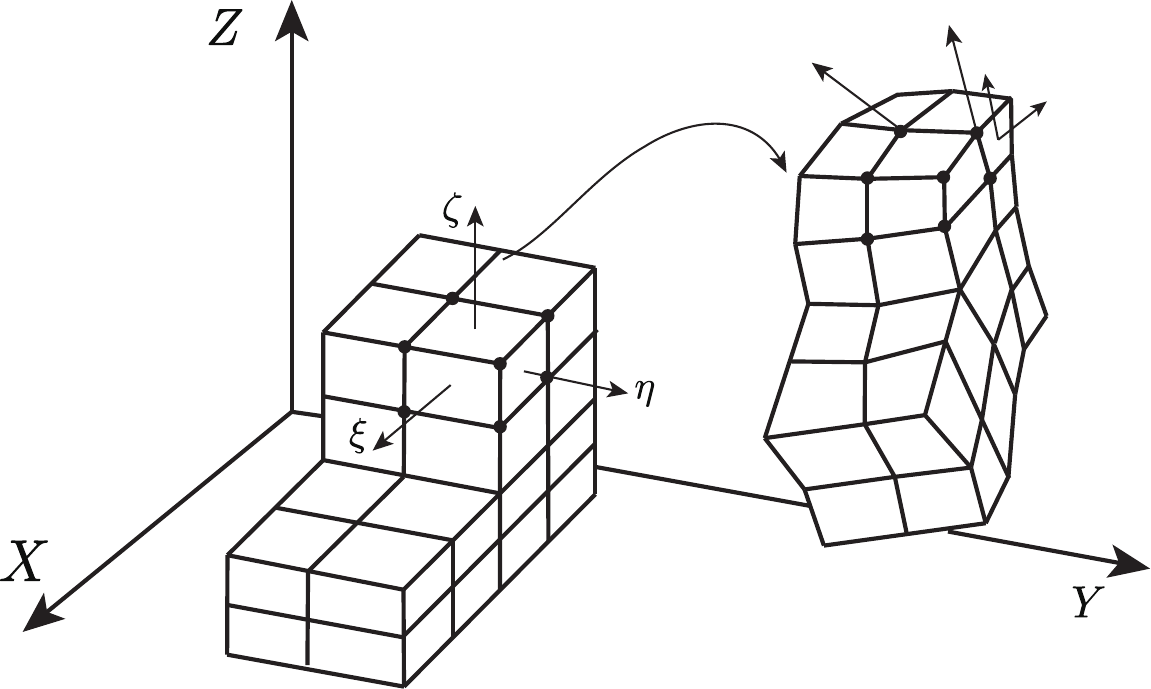}
    \caption{Discretization in the finite element method.}
    \label{p11.s3.FEM}
\end{figure}

In the finite element method, the domain $\Omega$ is discretized into a series of finite elements, as shown in Fig.\ref{p11.s3.FEM}.
Then, the displacement field within each element can be approximated by by ansatz functions $N_I$ as
\begin{equation}
    \bm{u}(\bm{X}) = \sum_{I=1}^{n_{node}} N_I(\bm{\xi})\bm{u}_I = \bm{N}^T\bm{U},
\end{equation}
where $n_{node}$ is the number of nodes for each element, $\bm{u}_I$ is the nodal displacement vector,
$N_I(\bm{\xi})$ is the shape function associated with node $I$, $\bm{\xi}$ is the local coordinate.
Besides, $\bm{N}$ is the shape function matrix and $\bm{U}$ is a vector for displacements associated with one element.

\subsection{Discrete formualtion}
\label{p11.s4.1}
As discussed before,
the strain energy density function for the third medium $\Omega_m$ is derived from the strain energy density function in the hyperelastic body $\Omega_s$.
The potential energy and it's linearization for the hyperelastic body $\Omega_s$ can be obtained from the third medium $\Omega_m$,
see Eqs.\eqref{p11.s3.1.W2} and \eqref{p11.s3.2.W2}.
So the strain energy density function $\Psi(\bm{F},\nabla\bm{F})$  for the third medium $\Omega_m$ is considered.
As discussed before, the second variation of the potential energy yields
\begin{equation}
    \label{p11.s4.1.w2}
    \delta^2 W^{TM} = \int_{\Omega_m}\left(
            \delta\bm{F}:\mathbb{C}:\delta\bm{F}
        +\delta\nabla\bm{F}\vdots\mathbb{A}:\delta\bm{F}
        +\delta\bm{F}:\mathbb{A}\vdots\delta\nabla\bm{F}
        +\delta\nabla \bm{F}\vdots\mathbb{B}\vdots\delta\nabla\bm{F}
        \right)\ud\Omega.
\end{equation}

In the framework of third medium contact, it is necessary to calculate the deformation gradient $\bm{F}$ and the gradient of the deformation gradient $\nabla\bm{F}$.
To facilitate matrix operations, we rewrite the variation of the deformation gradient $\bm{F}$ as a vector 
\begin{equation}
	\begin{aligned}
		\delta\hat{\bm{F}} = \begin{bmatrix}
			\frac{\partial\delta u_x}{\partial X} & \frac{\partial\delta u_y}{\partial X}& \frac{\partial\delta u_z}{\partial X} & 
			\frac{\partial\delta u_x}{\partial Y} & \frac{\partial\delta u_y}{\partial Y}& \frac{\partial\delta u_z}{\partial Y} &
            \frac{\partial\delta u_x}{\partial Z} & \frac{\partial\delta u_y}{\partial Z}& \frac{\partial\delta u_z}{\partial Z}
		\end{bmatrix}^T
		= \mathcal{L}_1\cdot
		\begin{bmatrix}
			\delta u_x \\ \delta u_y  \\ \delta u_z
		\end{bmatrix} = \bm{B}_1\delta\bm{U}.
	\end{aligned}
\end{equation}
Here, $u_x,u_y,u_z$ are the displacement components in $x,y,z$ directions, respectively,
and $\mathcal{L}_1$ is the first order differential operator
\begin{equation}
    \mathcal{L}_1 = \begin{bmatrix}
			\frac{\partial}{\partial X} & 0 & 0& \frac{\partial}{\partial Y} & 0 & 0 & \frac{\partial}{\partial Z} & 0 & 0 \\
			0 & \frac{\partial}{\partial X} & 0& 0 & \frac{\partial}{\partial Y} &0  & 0 & \frac{\partial}{\partial Z} &0 \\
            0 & 0 & \frac{\partial}{\partial X}&  0 & 0 & \frac{\partial}{\partial Y}& 0 & 0 & \frac{\partial}{\partial Z}
		\end{bmatrix}^T.
\end{equation}

In addition, the variation of the gradient of the deformation gradient $\nabla\bm{F}$ can also be arranged into a vector as
\begin{equation}
    \label{p11.s4.1.nablaF}
	\widehat{\nabla\delta\bm{F}}  = 
	\begin{bmatrix}
		\frac{\partial^2\delta u_x}{\partial X\partial X} & 
		\frac{\partial^2\delta u_y}{\partial X\partial X} &
		\frac{\partial^2\delta u_z}{\partial X\partial X} & 
        \frac{\partial^2\delta u_x}{\partial Y\partial X} & 
		\cdots & 
		\frac{\partial^2\delta u_x}{\partial Z\partial Z} & 
		\frac{\partial^2\delta u_y}{\partial Z\partial Z} & 
		\frac{\partial^2\delta u_z}{\partial Z\partial Z}
	\end{bmatrix}^T = \mathcal{L}_2\cdot
		\begin{bmatrix}
			\delta u_x \\ \delta u_y  \\ \delta u_z
		\end{bmatrix} = \bm{B}_2\delta\bm{U}.
\end{equation}
Here, $\mathcal{L}_2$ is the second order differential operator
\begin{equation}
    \mathcal{L}_2 = \begin{bmatrix}
			\frac{\partial^2}{\partial X\partial X} & 0 & 0& \frac{\partial^2}{\partial Y\partial X} & \cdots  & \frac{\partial^2}{\partial Z\partial Z} & 0 & 0 \\
			0 & \frac{\partial^2}{\partial X\partial X} & 0& 0 & \cdots   & 0 & \frac{\partial^2}{\partial Z\partial Z} &0 \\
            0 & 0 & \frac{\partial^2}{\partial X\partial X}&  0 & \cdots & 0 & 0 & \frac{\partial^2}{\partial Z\partial Z}
		\end{bmatrix}^T.
\end{equation}
The formulation of the second order differential of the shape functions will be discussed in the next section.

Based on the above definitions, the matrix formualtion of the variation of the potential energy (see Eq.\eqref{p11.s3.2.W1}) has the form as 
\begin{equation}
	\delta W^{TM} = \delta\bm{U}^T\int_{\Omega_m} \left(\bm{B}_1^T\hat{\bm{P}}+\bm{B}_2^T\hat{\bm{T}}\right) \ud\Omega,
\end{equation} 
where 
\begin{equation}
    \hat{\bm{P}} = 
    \begin{bmatrix}
        P_{11} & P_{21} & P_{31} & P_{12} & P_{22} & P_{32} & P_{13} & P_{23} & P_{33}
    \end{bmatrix}^T,
\end{equation}
and 
\begin{equation}
    \hat{\bm{T}} = 
    \begin{bmatrix}
        T_{111} & T_{211} & T_{311} & T_{121} & \cdots & T_{333}
    \end{bmatrix}^T.
\end{equation}

The second variation of the potential energy yields
\begin{equation}
    \label{p11.s4.1.ddw}
	\delta^2 W^{TM} = \delta\bm{U}^T\left[
		\int_{\Omega_m}
	\left(
		\bm{B}_1^T\hat{\mathbb{C}}\bm{B}_1+\bm{B}_2^T\hat{\mathbb{A}}\bm{B}_1+\bm{B}_1^T\hat{\mathbb{A}}^T\bm{B}_2+\bm{B}_2^T\hat{\mathbb{B}}\bm{B}_2
	\right)\ud\Omega
	\right]\delta\bm{U},
\end{equation}
where 
the matrices $\hat{\mathbb{C}},\hat{\mathbb{A}},\hat{\mathbb{B}}$ are square matrices with the components of the fourth-order tensor $\mathbb{C}$, 
the fifth-order tensor $\mathbb{A}$ and the sixth-order tensor $\mathbb{B}$.
The components of the matrix of the high-order tensors are given in the Appendix\ref{p11.appendix}

The tangent matrix on the element level can be obtained
\begin{equation}
    \label{p11.s4.1.Ke}
    \bm{K}_E = \int_{\Omega_m}
	\left(
		\bm{B}_1^T\hat{\mathbb{C}}\bm{B}_1+\bm{B}_2^T\hat{\mathbb{A}}\bm{B}_1+\bm{B}_1^T\hat{\mathbb{A}}^T\bm{B}_2+\bm{B}_2^T\hat{\mathbb{B}}\bm{B}_2
	\right)\ud\Omega,\quad \text{for }\bm{x}\in\Omega_m.
\end{equation}
For the hyperelastic bodies $\Omega_s$, the tangent stiffness matrix can be obtained from Eq.\eqref{p11.s4.1.Ke} directly 
\begin{equation}
    \label{p11.s4.1.Ke2}
    \bm{K}_E = \int_{\Omega_s}
	\bm{B}_1^T\hat{\mathbb{C}}\bm{B}_1\ud\Omega,\quad \text{for }\bm{x}\in\Omega_s.
\end{equation}

\subsection{Second order differential of shape functions}
\label{p11.s4.2}
The second order differential of the shape functions in Eq.\eqref{p11.s4.1.nablaF}needs the to calculation of the gradient of the deformation gradient $\nabla\bm{F}$.
Then second-order elements (for example, second-order serendipity element Q2S) 
should be used in the finite element method.
The gradient of shape functions with respect to the initial configuration can be obtained by
\begin{equation}
    \frac{\partial N_I}{\partial X_i} = \frac{\partial\xi_k}{\partial X_i}\frac{\partial N_I}{\partial\xi_k} = 
    \left(\bm{G}\right)_{ik}^{-1}\frac{\partial N_I}{\partial \xi_k}.
\end{equation}
Here we assume that we use Einstein's summation convention, which means that the same subindex represents the summation.
$\bm{G}$ is the Jacobian matrix between the local coordinate $\bm{\xi}$ and the initial coordinate $\bm{X}$:
\begin{equation}
    \label{Jacobi}
    \left.G_{ik}\right|_{E} = \left.\frac{\partial X_k}{\partial \xi_i}\right|_{E}.
\end{equation}
The inverse of the Jacobian matrix $\bm{G}$ can be calculated by 
\begin{equation}
    \label{p11.s4.2.inverseG}
    \bm{G}^{-1} = \frac{1}{\det(\bm{G})}\bm{G}^* = \frac{1}{|\bm{G}|}\bm{G}^*,
\end{equation}
where $\bm{G}^*$ is the adjoint matrix of $\bm{G}$
\begin{equation}
    \label{J1}
    \bm{G}^*=
    \begin{bmatrix}
        \dfrac{\partial Y}{\partial\eta}\dfrac{\partial Z}{\partial\zeta}-\dfrac{\partial Y}{\partial\zeta}\dfrac{\partial Z}{\partial\eta} & \dfrac{\partial Y}{\partial\zeta}\dfrac{\partial Z}{\partial\xi}-\dfrac{\partial Y}{\partial\xi}\dfrac{\partial Z}{\partial\zeta} & \dfrac{\partial Y}{\partial\xi}\dfrac{\partial Z}{\partial\eta}-\dfrac{\partial Y}{\partial\eta}\dfrac{\partial Z}{\partial\xi}\\
        \dfrac{\partial X}{\partial\zeta}\dfrac{\partial Z}{\partial\eta}-\dfrac{\partial X}{\partial\eta}\dfrac{\partial Z}{\partial\zeta} & \dfrac{\partial X}{\partial\xi}\dfrac{\partial Z}{\partial\zeta}-\dfrac{\partial X}{\partial\zeta}\dfrac{\partial Z}{\partial\xi} & \dfrac{\partial X}{\partial\eta}\dfrac{\partial Z}{\partial\xi}-\dfrac{\partial X}{\partial\xi}\dfrac{\partial Z}{\partial\eta}\\
        \dfrac{\partial X}{\partial\eta}\dfrac{\partial Y}{\partial\zeta}-\dfrac{\partial X}{\partial\zeta}\dfrac{\partial Y}{\partial\eta} & \dfrac{\partial X}{\partial\zeta}\dfrac{\partial Y}{\partial\xi}-\dfrac{\partial X}{\partial\xi}\dfrac{\partial Y}{\partial\zeta} & \dfrac{\partial X}{\partial\xi}\dfrac{\partial Y}{\partial\eta}-\dfrac{\partial X}{\partial\eta}\dfrac{\partial Y}{\partial\xi}
    \end{bmatrix},
\end{equation}
and
\begin{equation}
    \label{J2}
    \det(\bm{G}) = |\bm{G}| = \frac{\partial X}{\partial\xi}\frac{\partial Y}{\partial\eta}\frac{\partial Z}{\partial\zeta}+\frac{\partial X}{\partial\zeta}\frac{\partial Y}{\partial\xi}\frac{\partial Z}{\partial\eta}+\frac{\partial X}{\partial\eta}\frac{\partial Y}{\partial\zeta}\frac{\partial Z}{\partial\xi}
        - \frac{\partial X}{\partial\zeta}\frac{\partial Y}{\partial\eta}\frac{\partial Z}{\partial\xi}-\frac{\partial X}{\partial\xi}\frac{\partial Y}{\partial\zeta}\frac{\partial Z}{\partial\eta}-\frac{\partial X}{\partial\eta}\frac{\partial Y}{\partial\xi}\frac{\partial Z}{\partial\zeta}.
\end{equation}

The second order derivatives of the shape function can be deduced as 
\begin{equation}
    \label{p11.s4.2.second1}
    \begin{aligned}
        \frac{\partial^2 N_I}{\partial X_i\partial X_j} &= \frac{\partial}{\partial X_j}\left(\frac{\partial N_I}{\partial X_i}\right) 
    =\frac{\partial}{\partial X_j}\left(\left(\bm{G}\right)_{ik}^{-1}\frac{\partial N_I}{\partial \xi_k}\right)
    =\left(\left(\bm{G}\right)_{ik}^{-1}\frac{\partial^2N_I}{\partial\xi_l\partial\xi_k}+\frac{\partial N_I}{\partial\xi_k}\frac{\partial\left(\bm{G}\right)_{ik}^{-1}}{\partial \xi_l}\right)\frac{\partial\xi_l}{\partial X_j}
    \end{aligned}
\end{equation}
Based on Eqs.\eqref{p11.s4.2.inverseG},\eqref{J1} and \eqref{J2}, 
the derivative of the inverse matrix $\frac{\partial\bm{G}_{ik}^{-1}}{\partial\xi_l}$ in Eq.\eqref{p11.s4.2.second1} 
follows as
\begin{equation}
    \label{p11.s4.2.dinverse}
    \frac{\partial \bm{G}^{-1}}{\partial \xi_l} = \frac{1}{|\bm{G}|}\frac{\partial \bm{G}^*}{\partial\xi_l}
    -\frac{1}{|\bm{G}|^2}\frac{\partial|\bm{G}|}{\partial \xi_l}\bm{G}^*.
\end{equation}
The items in Eq.\eqref{p11.s4.2.dinverse} can be specified by differentiating Eqs.\eqref{J1} and \eqref{J2} directly.
Finally, by substituting Eq.\eqref{p11.s4.2.dinverse} into Eq.\eqref{p11.s4.2.second1},
the second order derivatives of the shape functions with respect to the initial configuration can be calculated.
All items in $\bm{B}_2$ can be computed and the tangent matrix in Eq.\eqref{p11.s4.1.Ke} can be calculated using Gaussian quadrature.

\section{Numerical examples}
\label{p11.s5}

\subsection{Self-contact within a 3D box}
\label{p11.s5.1}
\begin{figure}[htbp]
    \centering
    \includegraphics[width=1.0\textwidth]{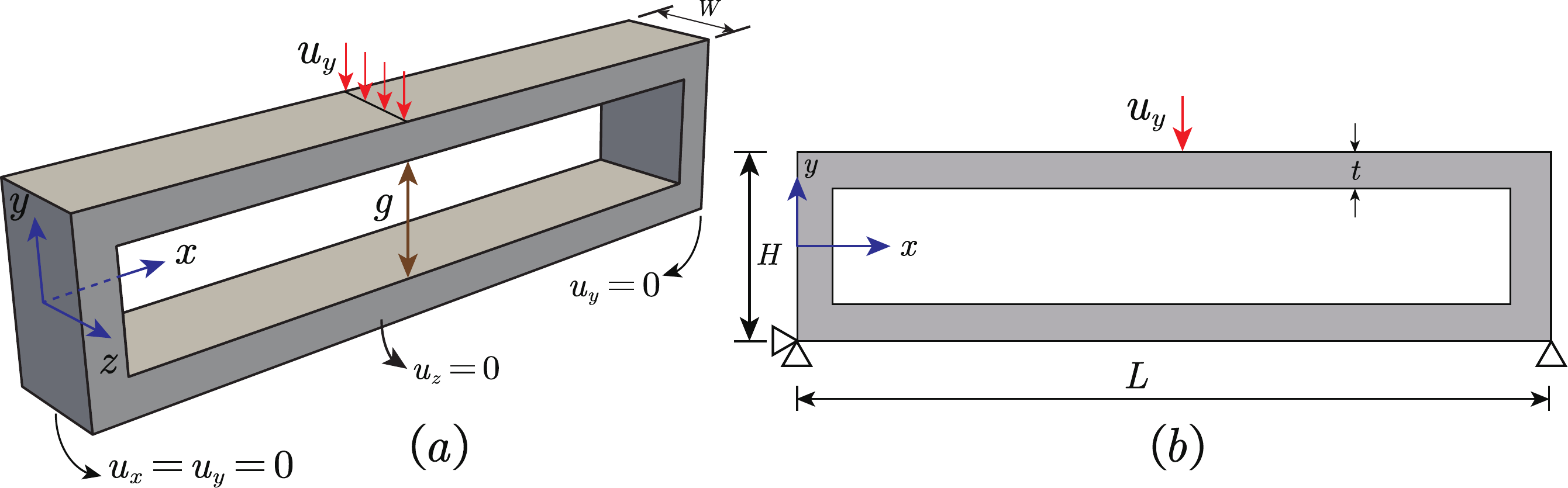}
    \caption{Self-contact within a 3D box, (a) geometry model and boundary condition, (b) front view of the 3D box.}
    \label{p11.s5.1.model}
\end{figure}

As a first example, we consider the self-contact within a 3D box, as shown in Fig.\ref{p11.s5.1.model}.
The box has a length of $L=2$ and a height of $H=0.5$, and the width is $W=0.3$.
Its thickness is $t=0.1$. 
The initial gap between the two plates (as shown in Fig.\ref{p11.s5.1.model} (a)) is $g_0=0.3$.
The boundary conditions are applied as shown in Fig.\ref{p11.s5.1.model} (a) with the fromt view in Fig.\ref{p11.s5.1.model} (b).
Note that the front and back faces ($z = 0.15$ and $z = -0.15$) of the box are also fixed as $u_z=0$.
A vertical displacement is prescribed as $u_y=-1.0$ at the top of the box (see Fig.\ref{p11.s5.1.model} (a)).

\begin{figure}[htbp]
    \centering
    \includegraphics[width=1.0\textwidth]{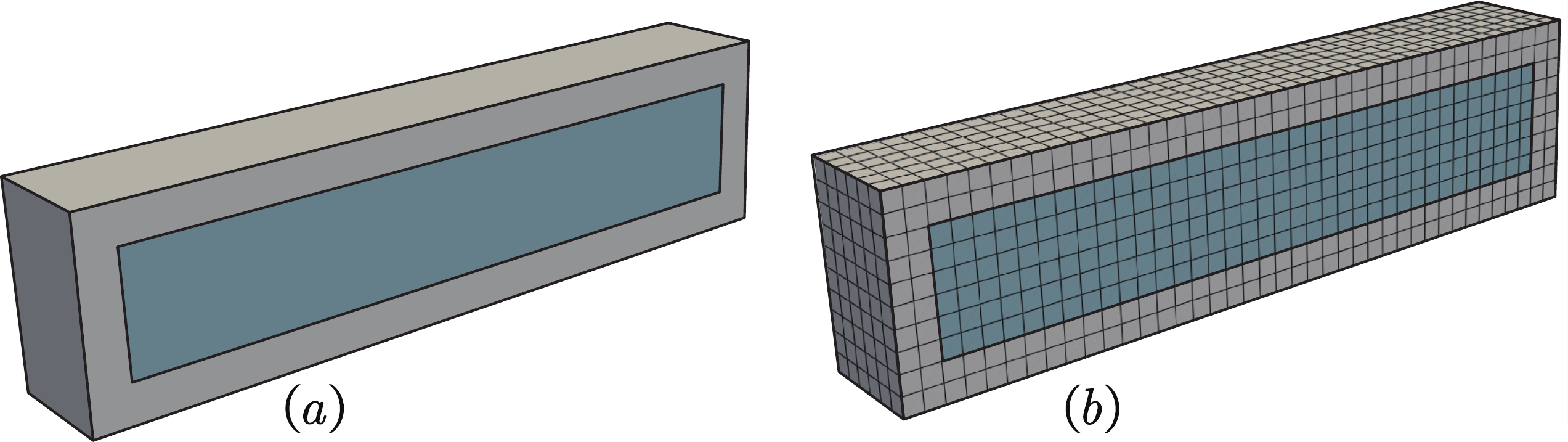}
    \caption{Self-contact within a 3D box, (a) geometry model and the third medium, (b) finite element mesh.}
    \label{p11.s5.1.mesh}
\end{figure}

The strain energy density function describing the box material is the Neo-Hookean model as given in Eq.\eqref{p11.s2.strainEnergy},
with the shear modulus $K=20$ and the bulk modulus $\mu=10$.
The second-order serendipity element (Q2S) is used for the finite element analysis.
The third medium fills the box completely, as depicted in Fig.\ref{p11.s5.1.mesh} (a).
The finite element mesh for the box and the third medium is shown in Fig.\ref{p11.s5.1.mesh} (b).

\begin{figure}[htbp]
    \centering
    \includegraphics[width=1.0\textwidth]{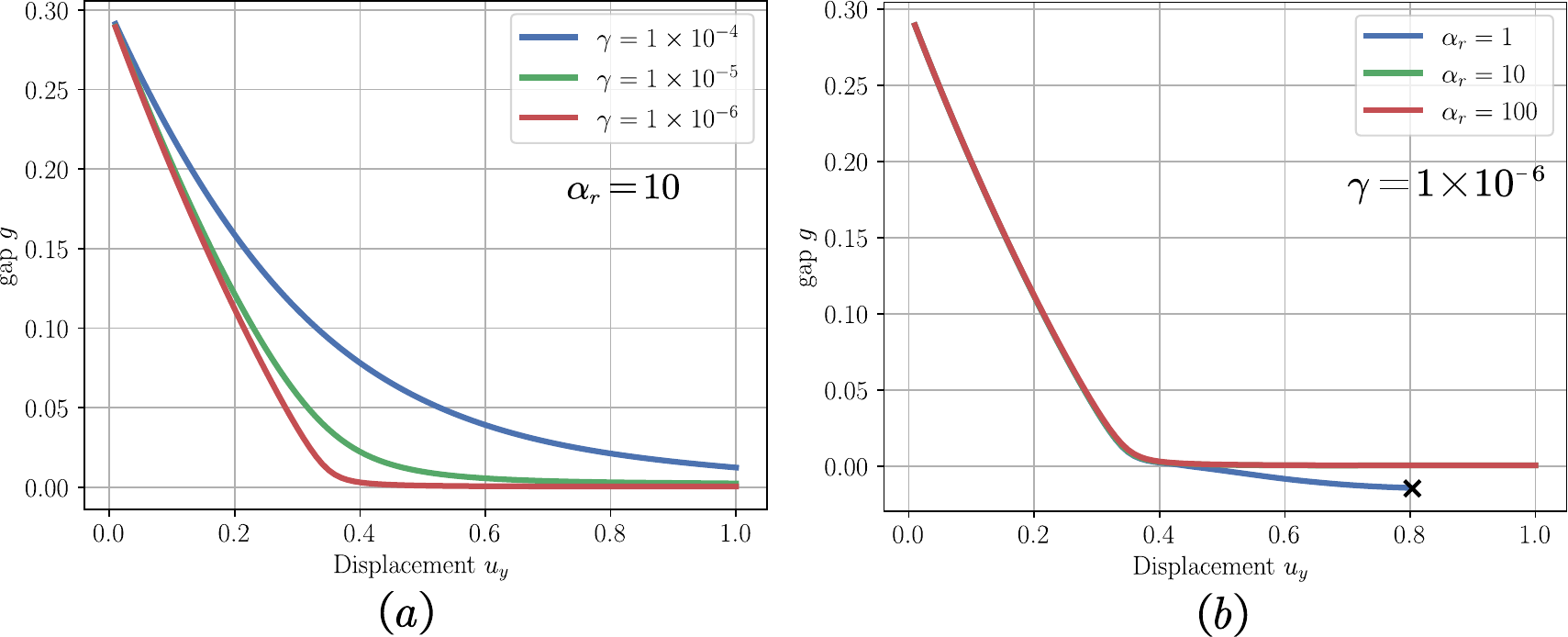}
    \caption{Gap $g$ between the surfaces of the upper and lower flange for different parameters, 
    (a) different $\gamma$ and $\alpha_r = 10$,
    (b) different $\alpha_r$ and $\gamma = 1\times 10^{-6}$.}
    \label{p11.s5.1.line}
\end{figure}

We investigate the effects of different regularization parameters on the simulation results.
The regularization term is selected as Eq.\eqref{p11.s2.3.fskew}.
Different parameters $\alpha_r$ and $\gamma$ are used to investigate their effects on the simulation results.
The gap $g$ (the distance between the midpoint of the upper surface and the midpoint of the lower surface) is investigated,
and the results are shown in Table \ref{p11.s5.1.t1}.
It can be seen that the result of $g$ is greatly related to the choice of the parameter $\gamma$.
A smaller $\gamma$ will result in a smaller gap value $g$. 
The gap $g$ for different parameters is also shown in Fig.\ref{p11.s5.1.line} for some specific parameters.
However, as $\gamma$ decreases, the third medium becomes softer and leads to element overlap which causes the calculation failure (see Fig.\ref{p11.s5.1.line} (b)). 
The stability can be improved by increasing the parameter $\alpha_r$ in the regularization term.
So in the next examples, we will choose a large parameter $\alpha_r$ to ensure the stability and convergence of the simulation.

\begin{table}[htbp]
\centering
\caption{Gap $g$ between the surfaces of the upper and lower flange for different parameters.}
\label{p11.s5.1.t1}
\begin{tabular}{cccccc}
\hline
\multicolumn{1}{c}{$\alpha_r$} & \multicolumn{1}{c}{$\gamma$} &  \multicolumn{1}{c}{$g$}  \\ \hline
\multirow{3}{*}{100}            & $1\times 10^{-4}$      &       1.2414E-02               \\
                               & $1\times 10^{-5}$      &       2.4393E-03               \\
                               & $1\times 10^{-6}$      &       5.1653E-04               \\ \hline
\multirow{3}{*}{10}           & $1\times 10^{-4}$      &       1.1135E-2              \\
                               & $1\times 10^{-5}$      &       2.4206E-03               \\
                               & $1\times 10^{-6}$      &       4.9783E-04               \\ \hline
\multirow{3}{*}{1}           & $1\times 10^{-4}$      &       1.0995E-02               \\
                               & $1\times 10^{-5}$      &       2.3138E-03               \\
                               & $1\times 10^{-6}$      &       Calculation failed                    \\ \hline
\end{tabular}
\end{table}

\begin{figure}[htbp]
    \centering
    \includegraphics[width=1.0\textwidth]{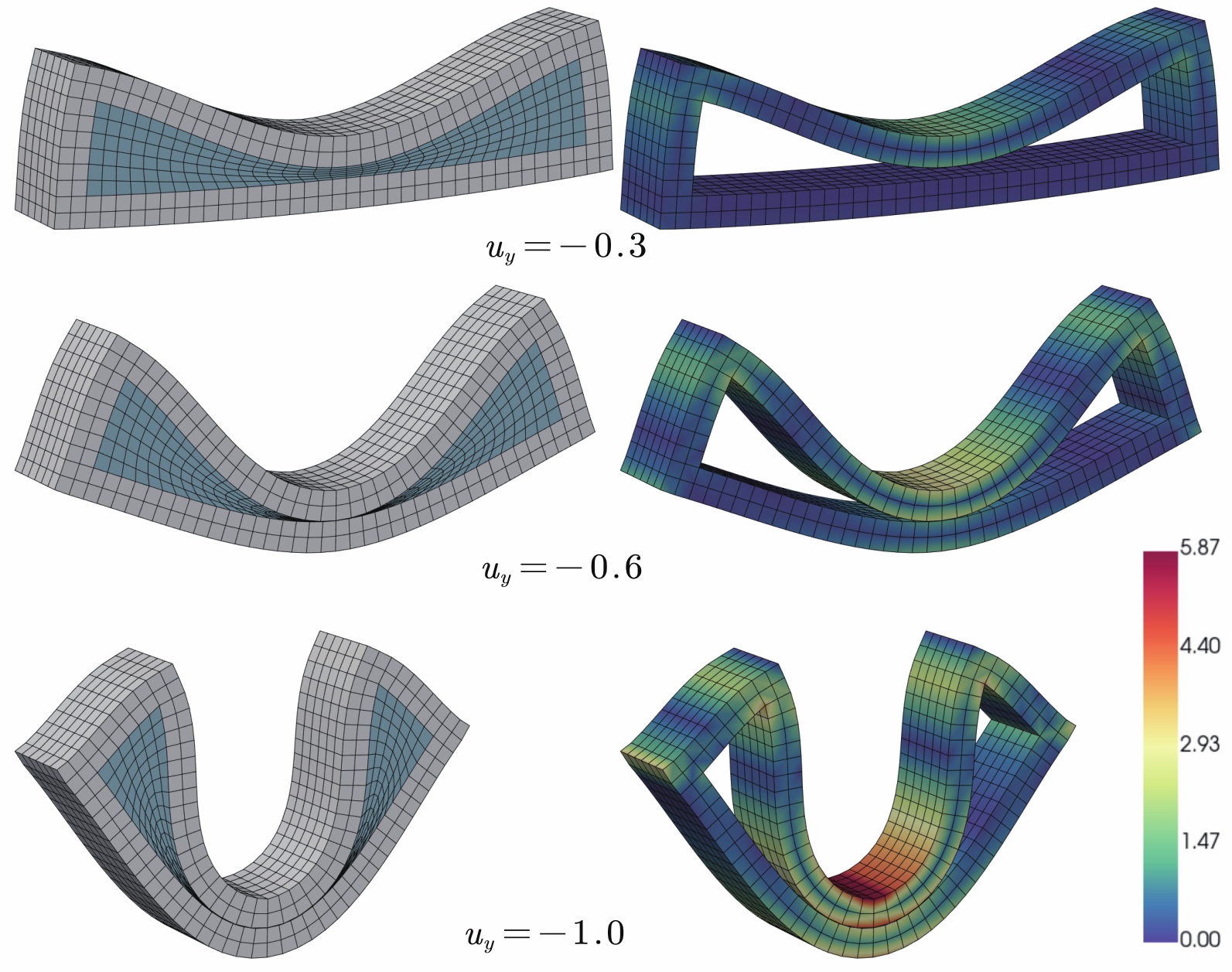}
    \caption{Deformed configuration and contour of von-Mises stress for the box under different displacement $u_y$.}
    \label{p11.s5.1.c}
\end{figure}

For parameters $\alpha_r=10$ and $\gamma=1\times 10^{-6}$,
the deformed configurations and the contours of von-Mises stress for different displacement $u_y$ are shown in Fig.\ref{p11.s5.1.c}.

\subsection{Self-contact caused by rotation of a 3D box}
\label{p11.s5.2}
\begin{figure}[htbp]
    \centering
    \includegraphics[width=1.0\textwidth]{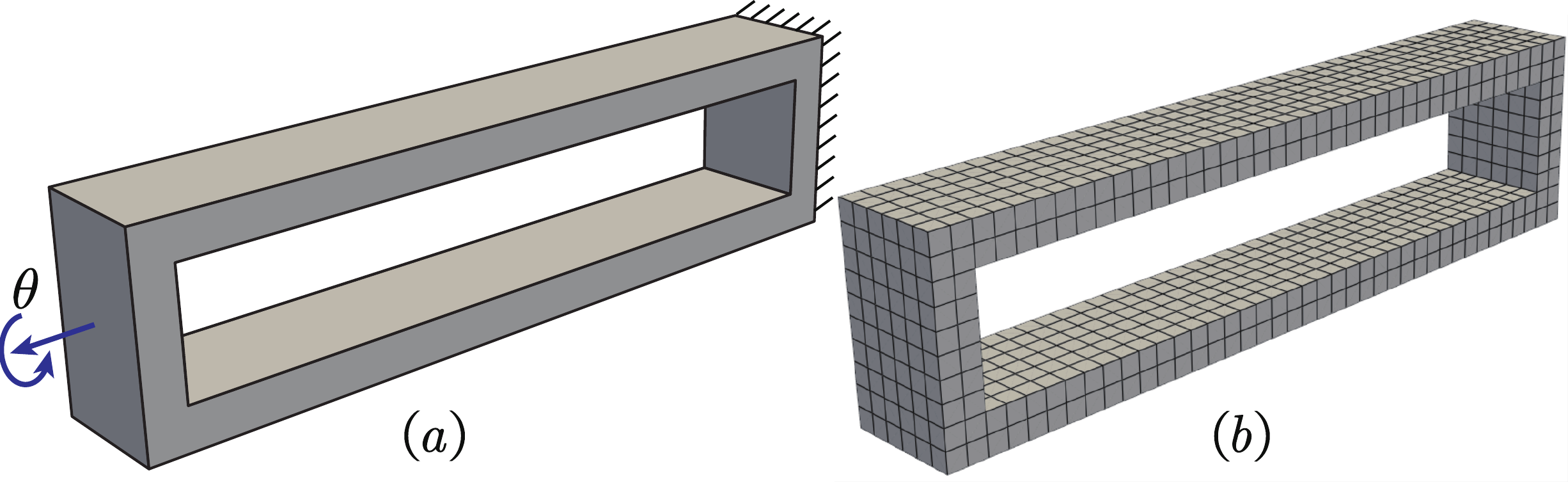}
    \caption{Self-contact within a 3D box, (a) geometry model and boundary condition, (b) FEM mesh of the solid domain.}
    \label{p11.s5.2.model}
\end{figure}

\begin{figure}[htbp]
    \centering
    \includegraphics[width=1.0\textwidth]{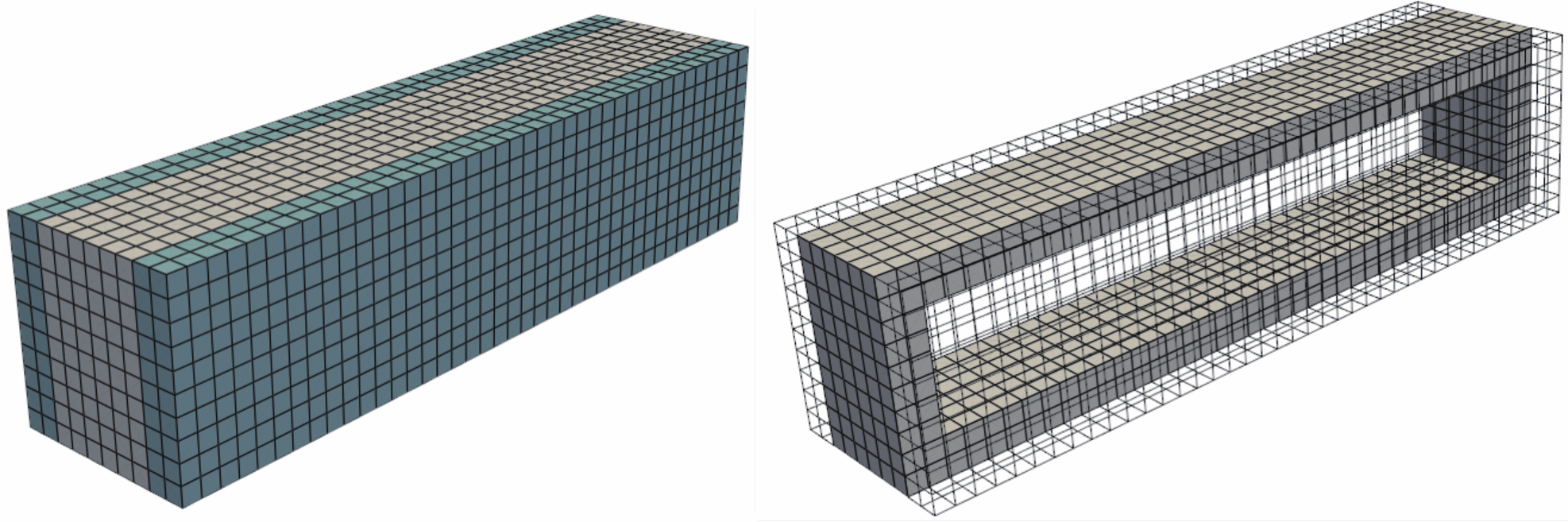}
    \caption{FEM mesh for the third medium of the 3D box.}
    \label{p11.s5.2.mesh}
\end{figure}

In this example, we consider the self-contact problem caused by the rotation of a 3D box, 
as shown in Fig.\ref{p11.s5.2.model}.
The geometry and material properties of the box are the same as those in Section \ref{p11.s5.1}.
The boundary conditions are applied as shown in Fig.\ref{p11.s5.2.model}.
The right side of the box is fixed, while a rotation angle $\theta$ is applied at the left side of the box.
Because of the potential contact of free surfaces, 
the third medium region should be appropriately expanded when defining it.
As shown in Fig.\ref{p11.s5.2.mesh}, the third medium region includes the cavity of the box and an additional surrounding region
(located on the front and back sides of the box, with a thickness of $0.1$).
This ensures that the contact is correctly recognized.

\begin{figure}[htbp]
    \centering
    \includegraphics[width=1.0\textwidth]{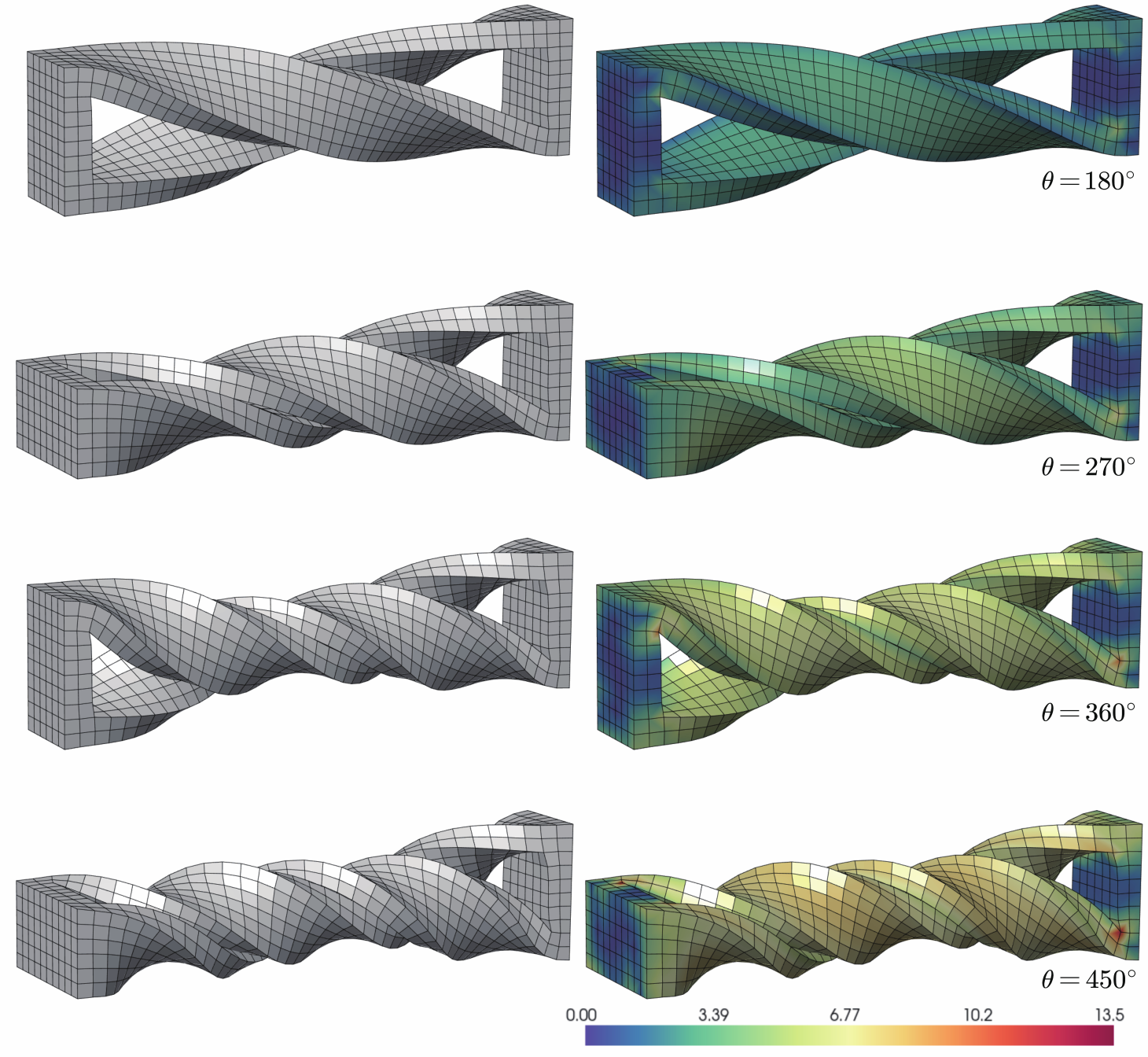}
    \caption{Deformed configuration and contour of von-Mises stress for the box under different rotation angle $\theta$.}
    \label{p11.s5.2.c}
\end{figure}

Considering the complexity of this problem, we set the parameters to be relatively large as $\alpha_r=10$ and $\gamma=1\times 10^{-4}$.
The rotation angle $\theta$ is increased incrementally from $0^\circ$ to $450^\circ$.
The Newton-Raphson algorithm is used to solve the problem with 125 steps ($\Delta\theta = 3.6^\circ$) with an average of 5 iterations perstep.
For this self-contact problem, if the classic contact algorithm is used, 
the search for the contact surface and the subsequent nonlinear iterations will be time-consuming and will need a robust strategy.
In the third medium framework, the contact problem is transformed into a standard boundary value problem,
making the simulation process straightforward and stable.
The deformed configurations and the contours of the von-Mises stress are shown in Fig.\ref{p11.s5.2.c} for different rotation angle $\theta$.

\subsection{Punch problem}
\label{p11.s5.3}
\begin{figure}[htbp]
    \centering
    \includegraphics[width=1.0\textwidth]{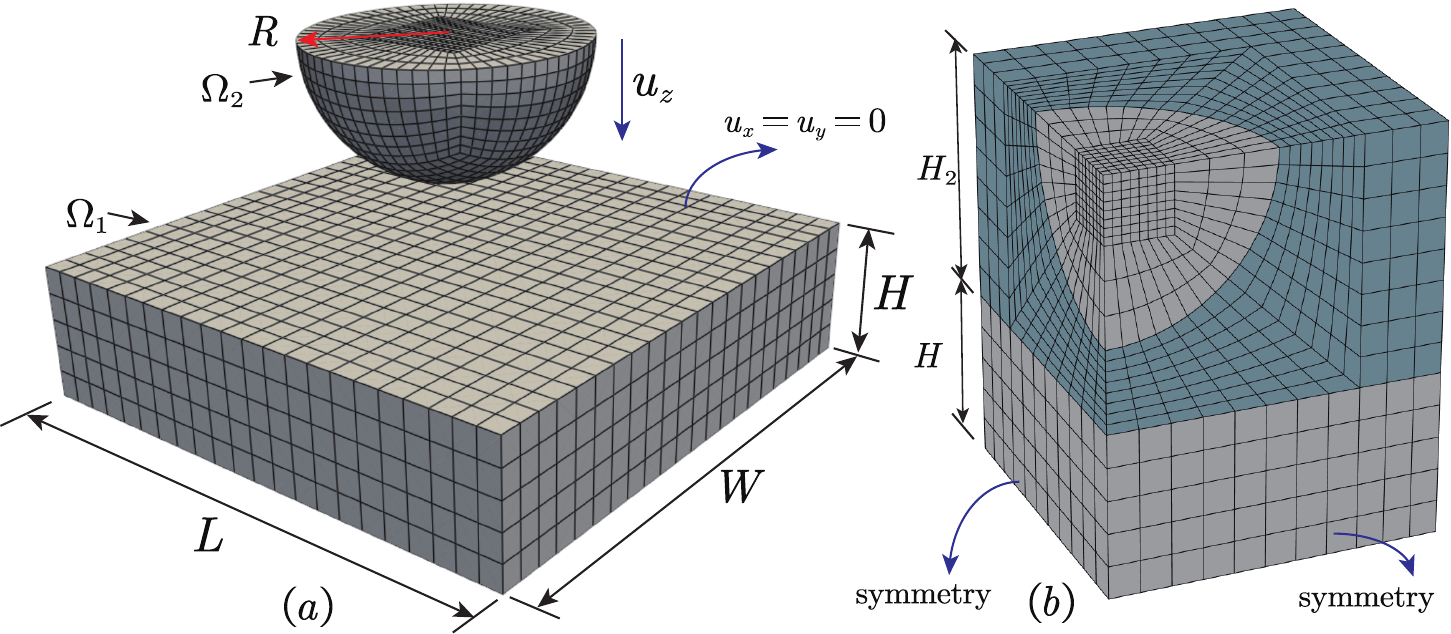}
    \caption{Punch problem, (a) geometry model and boundary condition, (b) finite element mesh and third medium.}
    \label{p11.s5.3.model}
\end{figure}

In this example, we consider the punch problem as shown in Fig.\ref{p11.s5.3.model}.
A hemispherical punch, $\Omega_2$, with a radius $R=1$ is used to indent a block $\Omega_1$ with length of $L=4.0$, 
height of $H=1.0$ and width of $W=4.0$.
A vertical displacement $u_z$ is prescribed at the top of the hemisphere $\Omega_2$.
The boundary conditions are applied as shown in Fig.\ref{p11.s5.3.model}.
The bottom surface of the block $\Omega_1$ is fixed on the $z$ direction.
Due to symmetry, only $1/4$ model is considered in the simulation.

The Neo-Hookean model as given in Eq.\eqref{p11.s2.strainEnergy} defines the strain energy density function for the two bodies.
The shear and bulk modulus for the block $\Omega_1$ are selected as $K_1=5/4$ and $\mu_1=5/14$,
while for the hemisphere punch $\Omega_2$, they are selected as $K_2=500/4$ and $\mu_2=500/14$.
The third medium completely fills the space between the two bodies, as shown in Fig.\ref{p11.s5.3.model} (b).
The initial gap between the two bodies is $g_0=0.5$ ($H_2 = 1.5$, see Fig.\ref{p11.s5.3.model}).
The parameters for the third medium are selected as $\gamma = 1\times 10^{-4}$ and $\alpha_r = 100$.

\begin{figure}[htbp]
    \centering
    \includegraphics[width=1.0\textwidth]{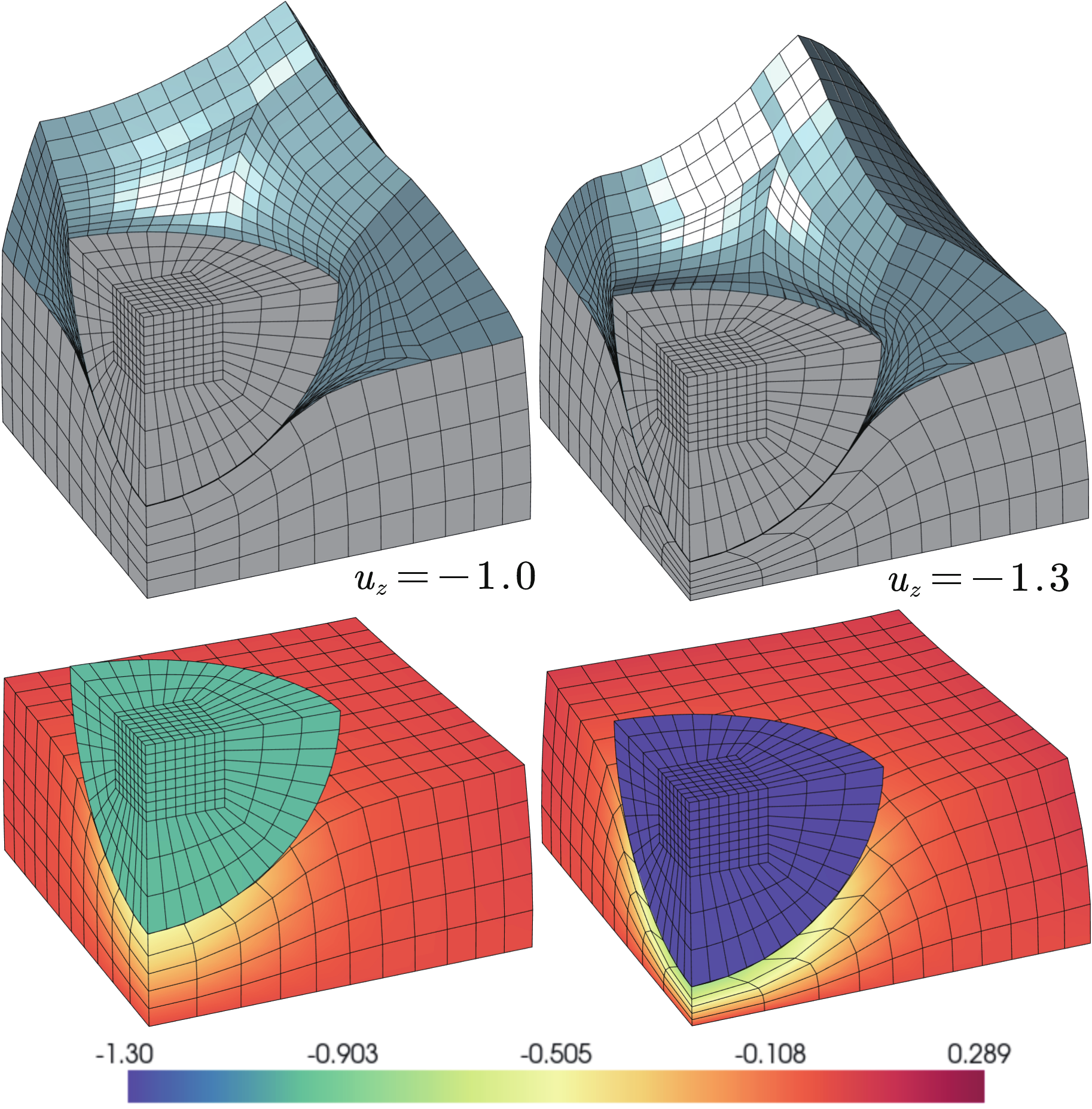}
    \caption{Deformed configurations under displacement load $u_z$ for the punch problem.}
    \label{p11.s5.3.d}
\end{figure}

\begin{figure}[htbp]
    \centering
    \includegraphics[width=1.0\textwidth]{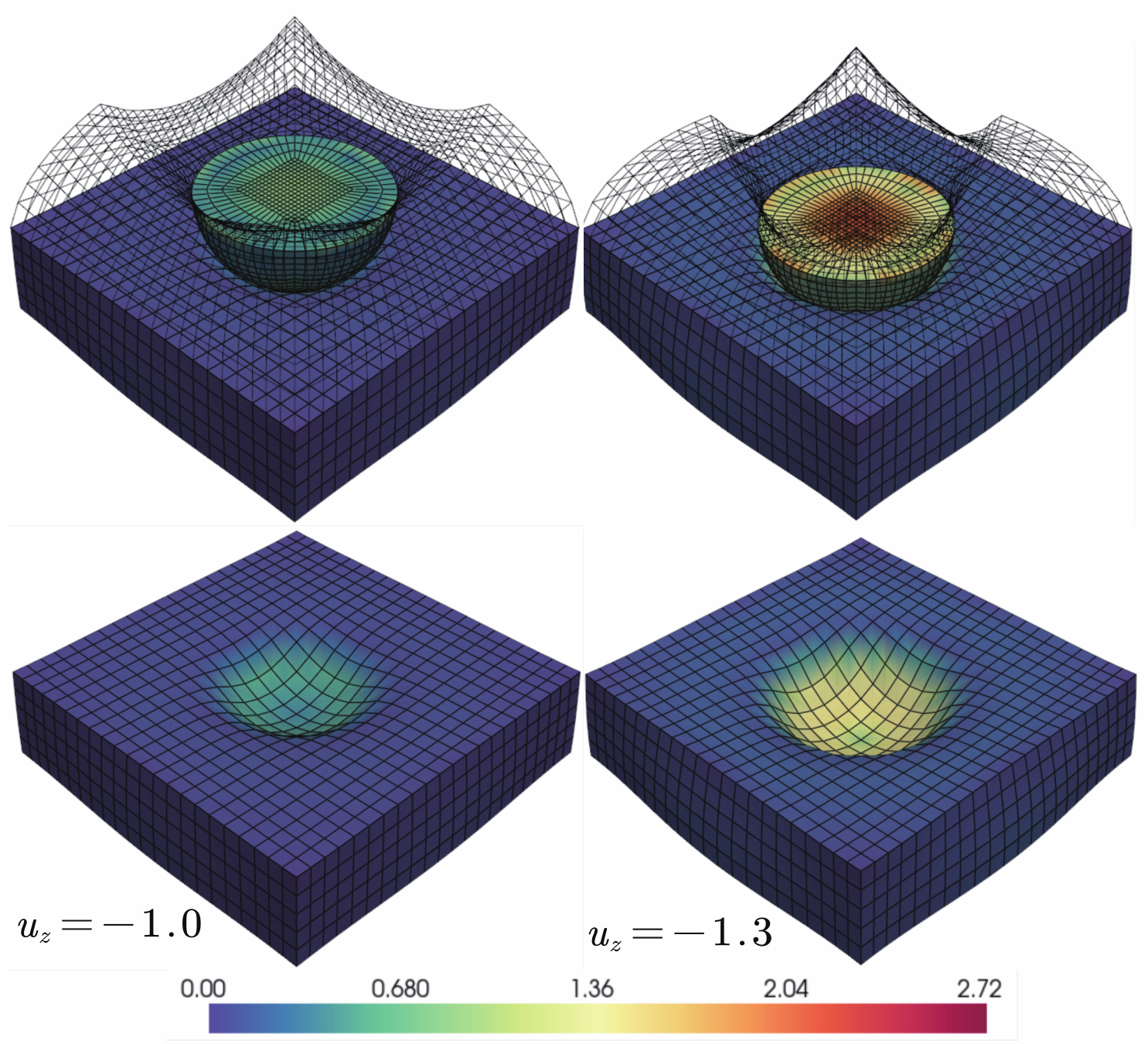}
    \caption{Von-Mises stress under displacement load $u_z$ for the punch problem.}
    \label{p11.s5.3.c}
\end{figure}

The prescribed vertical displacement is selected as $u_z=-1.4$ and applied at the top of the hemisphere $\Omega_2$.
The Newton-Raphson scheme is used with 140 steps (4 to 5 iterations per step).
The deformed configuration for different prescribed displacements $u_z$ are shown in Fig.\ref{p11.s5.3.d}.
The contours of von-Mises stress for the punch problem under different displacement load $u_z$ are shown in Fig.\ref{p11.s5.3.c}.

\subsection{Third medium contact for the pneumatically actuated system}
\label{p11.s5.4}
\begin{figure}[htbp]
    \centering
    \includegraphics[width=1.0\textwidth]{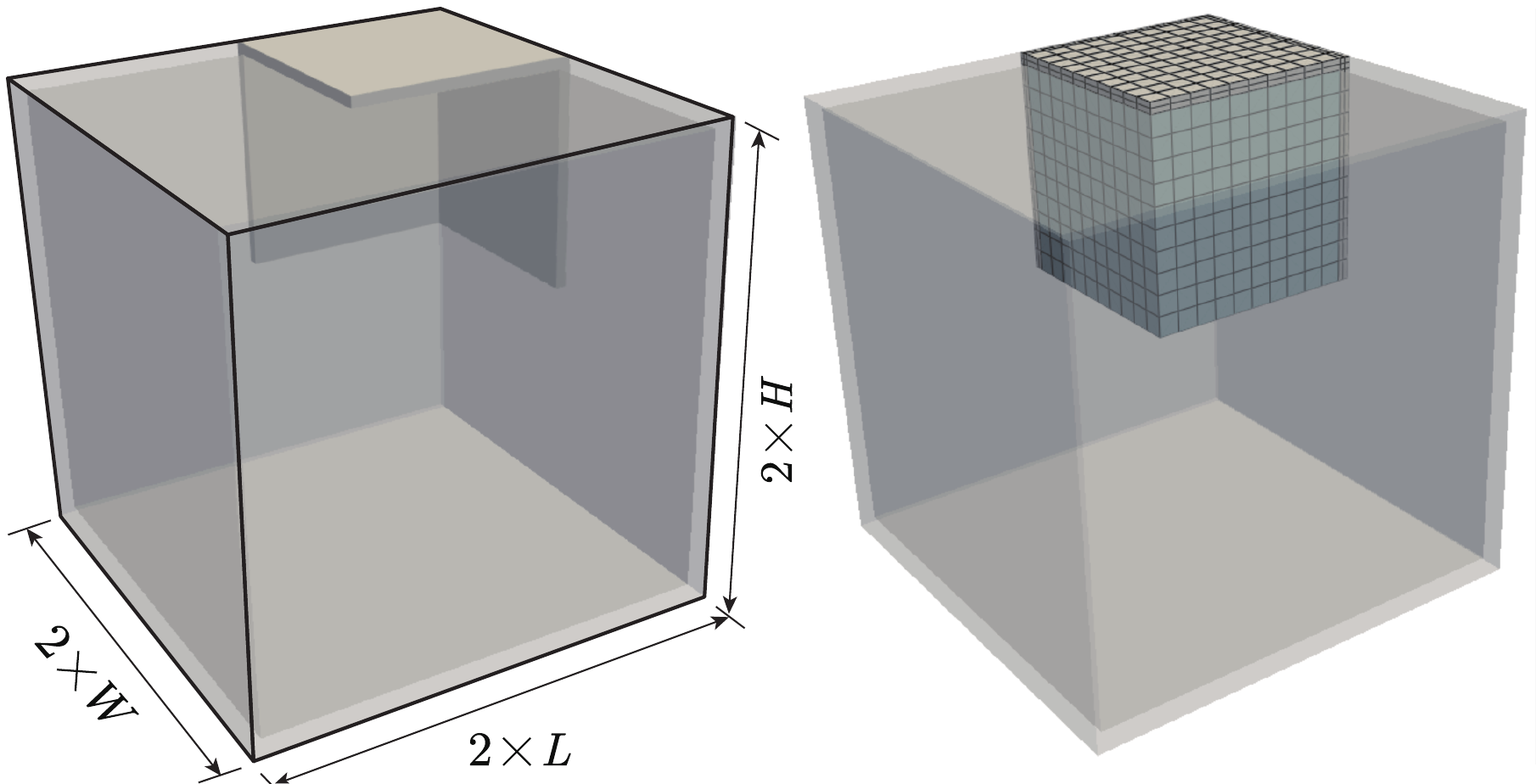}
    \caption{Pneumatically actuated box with third medium.}
    \label{p11.s5.4.model}
\end{figure}

As discussed before, the third medium contact framework can be applied to a pneumatically actuated system by introducing the pressure part in the strain energy density function of the third medium.
In this example, we consider a specific problem to test the performance of the third medium theory in a pneumatically actuated system.
As shown in Fig.\ref{p11.s5.4.model}, a cube-shaped box with a cavity is considered.
The box has a length of $2\times L=2.0$ and a height of $2\times H=2.0$, and a width of $2\times W=2.0$.
The thickness of the box is $t=0.5$.
Due to symmetry, only $1/8$ model is considered.

Again, the Neo-Hookean model as given in Eq.\eqref{p11.s2.strainEnergy} was used
with the shear modulus $K=20$ and the bulk modulus $\mu=10$.
The third medium completely fills the cavity of the box, as shown in Fig.\ref{p11.s5.4.model}.
The parameters for the third medium are selected as $\gamma = 1\times 10^{-5}$ and $\alpha_r = 100$.
The suction or pressure is applied using Eq.\eqref{p11.s2.2.PsiP}.
The finite element mesh for the box and the third medium is depicted in Fig.\ref{p11.s5.4.model}.

\begin{figure}[htbp]
    \centering
    \includegraphics[width=0.9\textwidth]{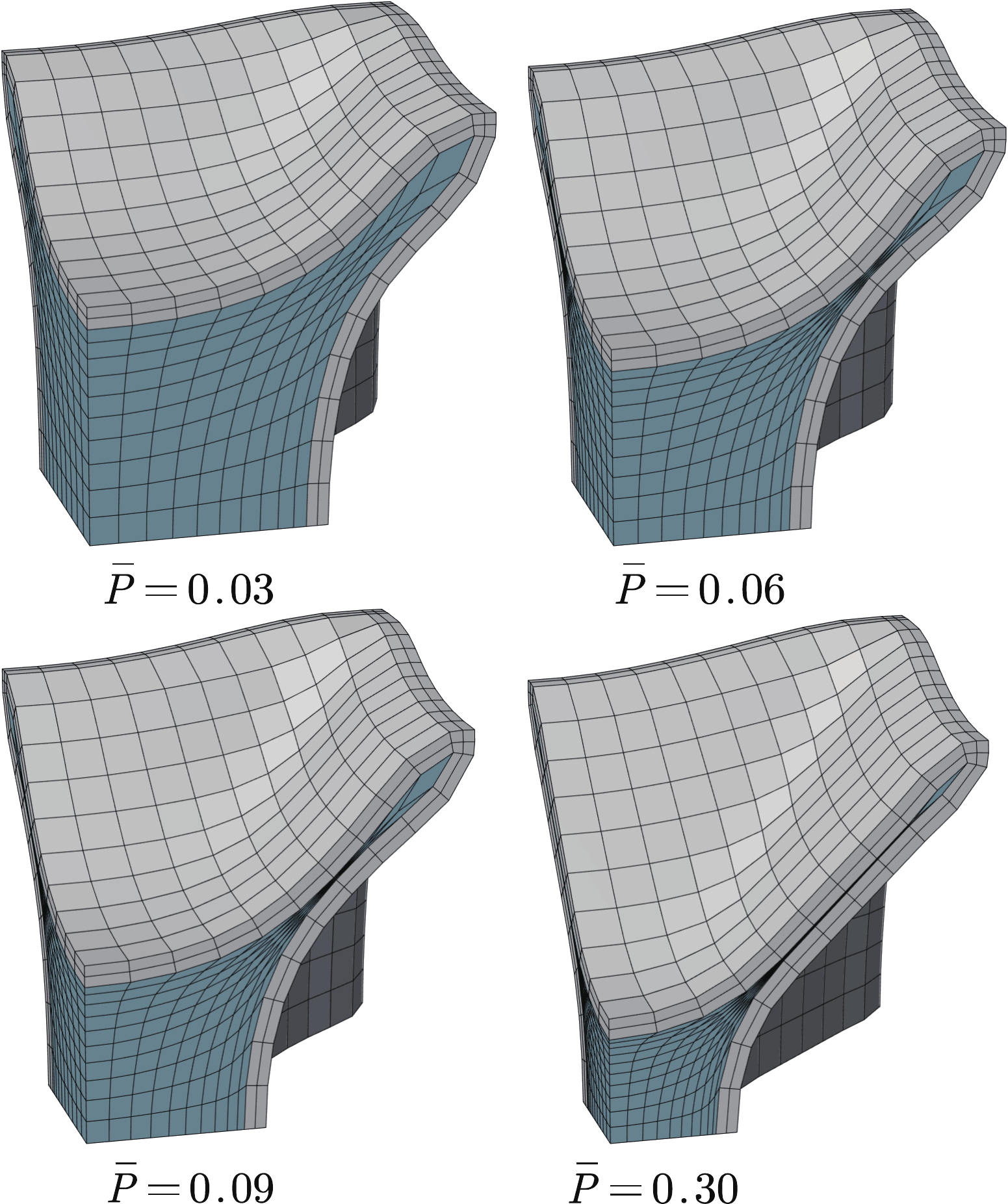}
    \caption{Deformed configuration of the pneumatically actuated box under suction, 1/8 model.}
    \label{p11.s5.4.def1}
\end{figure}

\begin{figure}[htbp]
    \centering
    \includegraphics[width=1.0\textwidth]{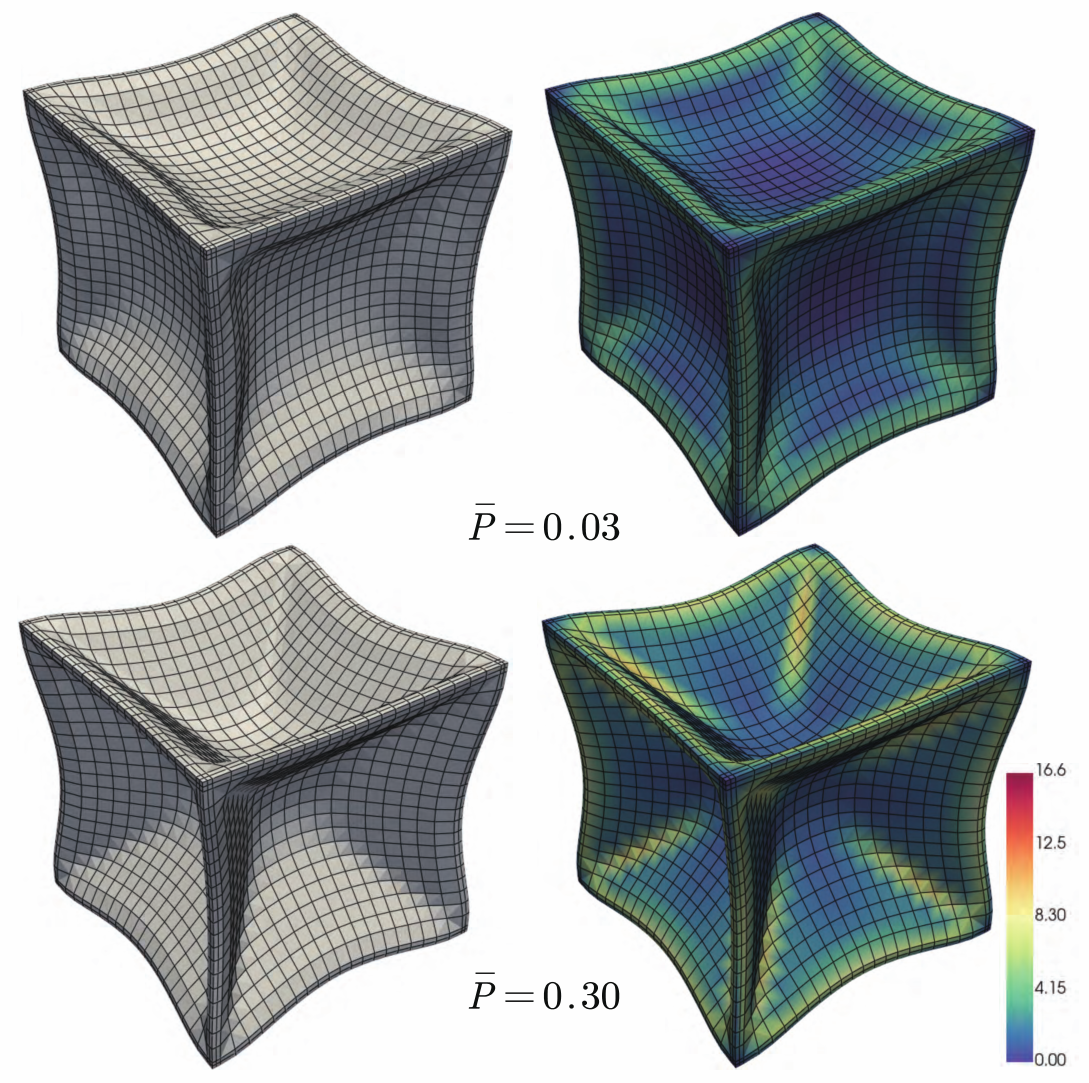}
    \caption{Deformed configuration and contour of von-Mises stress for the pneumatically actuated box under and suction, full model.}
    \label{p11.s5.4.def2}
\end{figure}

Firstly, the magnitude of the suction is selected as $\bar{P}=0.3$.
The Newton-Raphson scheme is employed with 100 steps (4 to 5 iterations per step).
Due to the presence of internal suction, the box deforms and contracts inward. 
The deformation under different internal suction conditions is shown in Fig.\ref{p11.s5.4.def1}. 
Observed that self-contact occurs when the suction is large ($\bar{P} = 0.3$). 
Therefore, this third-medium formualtion can not only apply pneumatic forces but automatically treats self-contact caused by pneumatic forces.
The contours of the von-Mises stress for the pneumatically actuated box under different magnitude of the suction $\bar{P}$ are visualized in Fig.\ref{p11.s5.4.def2}.

\begin{figure}[htbp]
    \centering
    \includegraphics[width=1.0\textwidth]{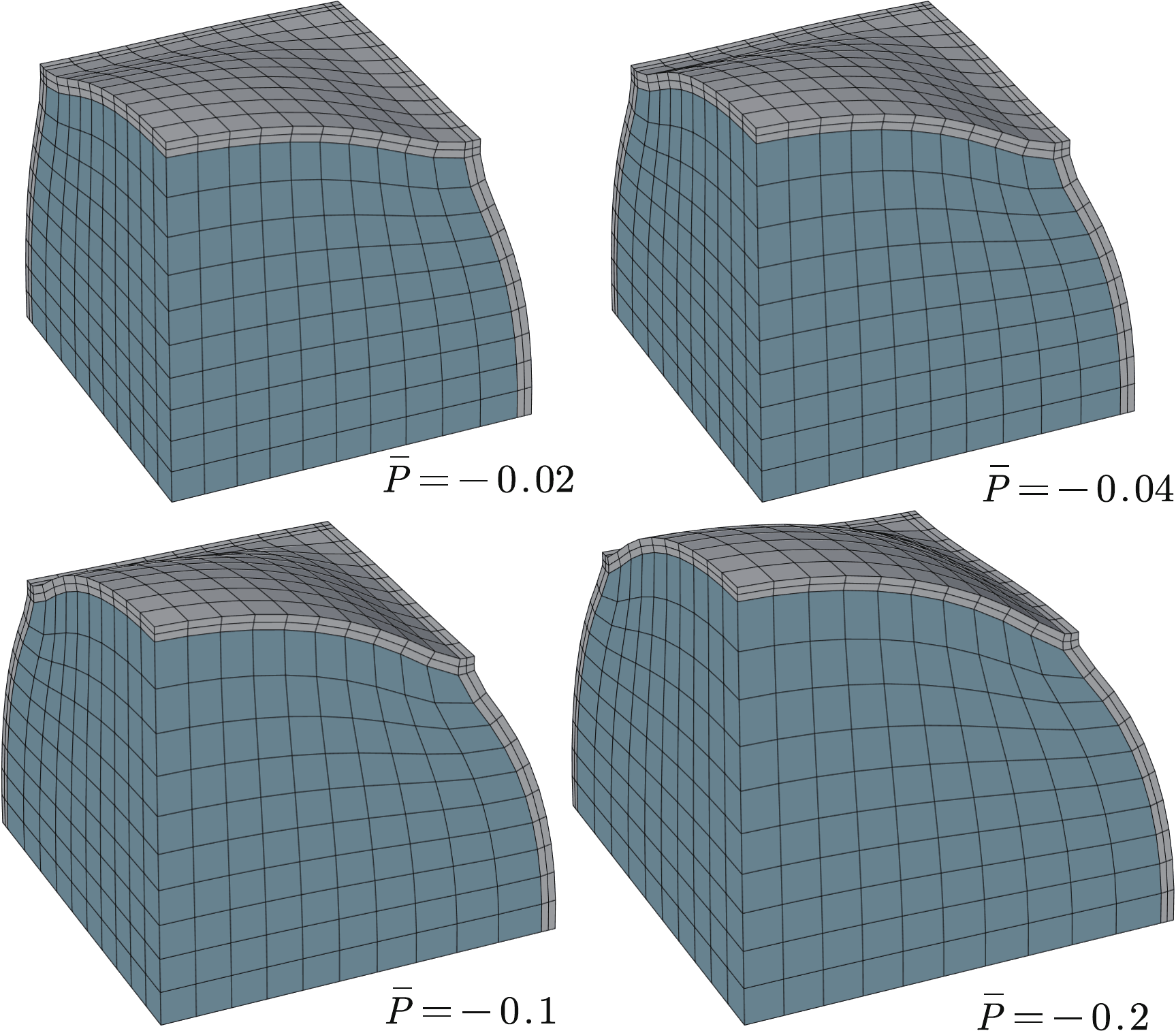}
    \caption{Deformed configuration of the pneumatically actuated box under pressure, 1/8 model.}
    \label{p11.s5.4.def3}
\end{figure}

\begin{figure}[htbp]
    \centering
    \includegraphics[width=1.0\textwidth]{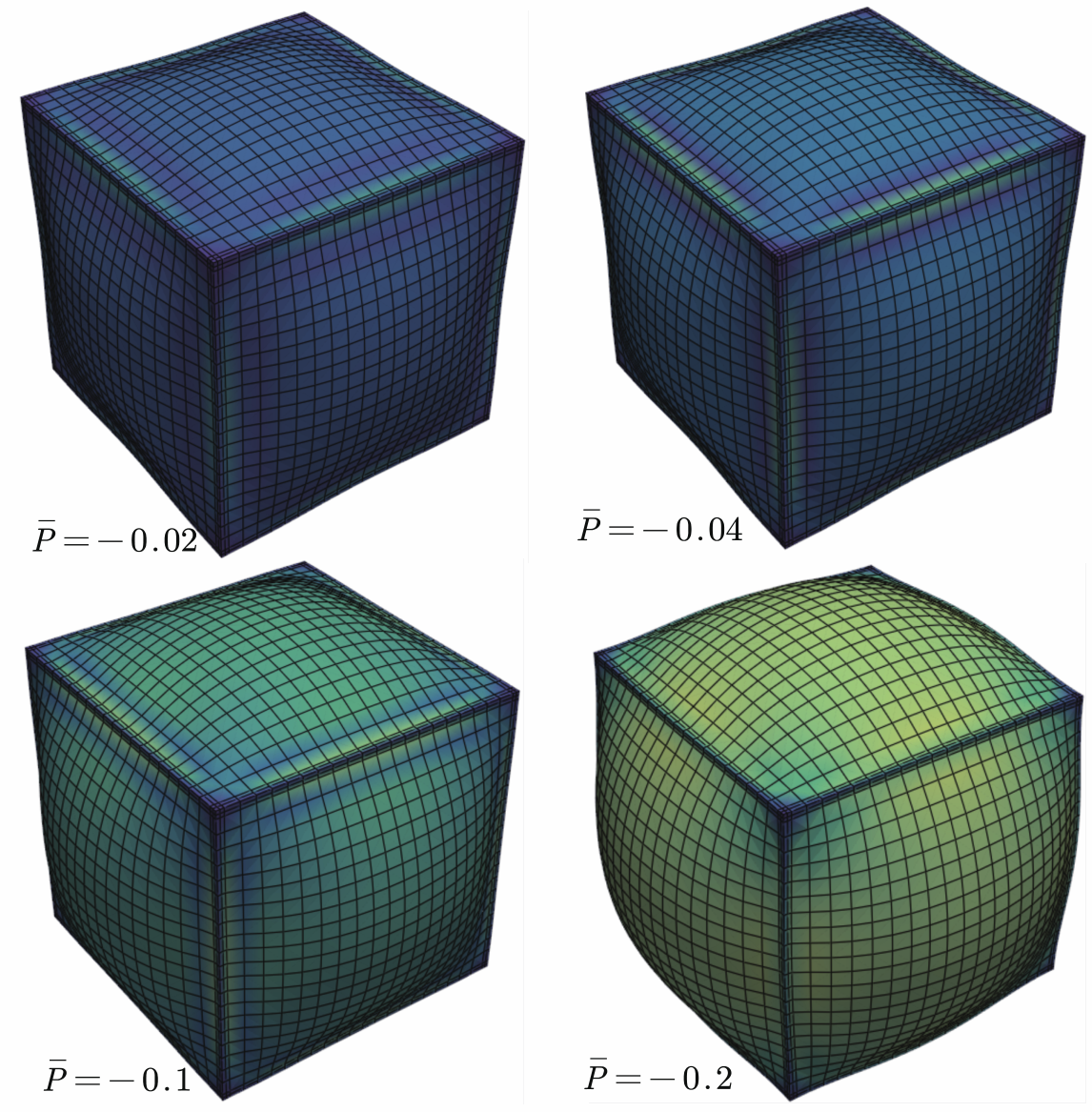}
    \caption{Contour of von-Mises stress for the pneumatically actuated box under and pressure, full model.}
    \label{p11.s5.4.def4}
\end{figure}

In the next loading cases, internal pressure is assumed and the magnitude of the pneumatic part is selected as $\bar{P}=-0.2$.
The parameters for the third medium are selected as $\gamma = 1\times 10^{-5}$ and $\alpha_r = 100$.
The deformations of the box under different internal pressure conditions are depicted in Fig.\ref{p11.s5.4.def3}.
It can be seen that by adding a pneumatic term to the third medium, the deformation caused by air pressure can be effectively simulated.
The contours of von-Mises stress for the pneumatically actuated box under different magnitude of the pressure $\bar{P}$ are shown in Fig.\ref{p11.s5.4.def4}.

\subsection{Pneumatic soft robot actuator}
\label{p11.s5.5}
\begin{figure}[htbp]
    \centering
    \includegraphics[width=1.0\textwidth]{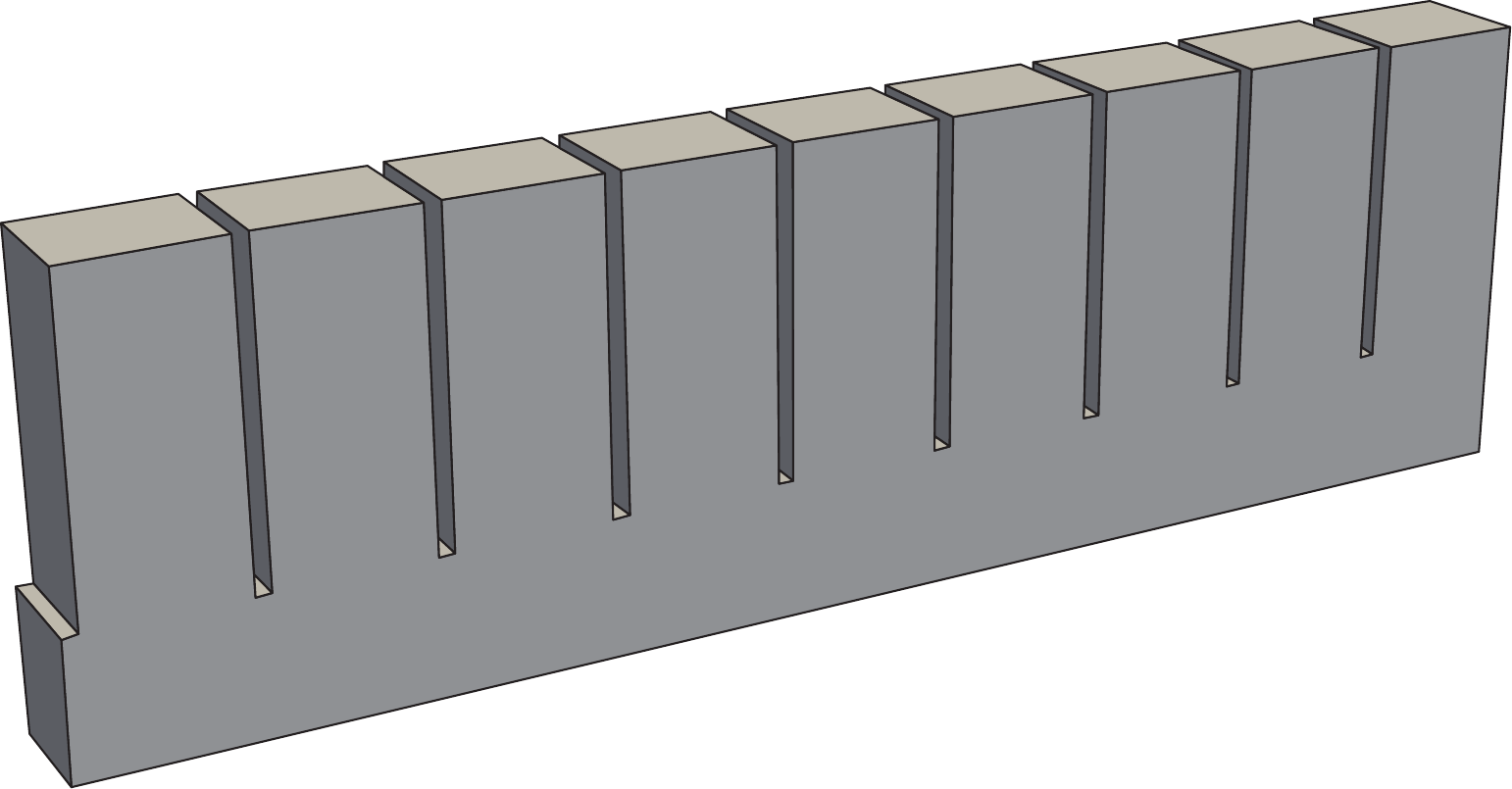}
    \caption{Geometry model for the pneumatically soft robot actuator.}
    \label{p11.s5.5.model}
\end{figure}

\begin{figure}[htbp]
    \centering
    \includegraphics[width=0.8\textwidth]{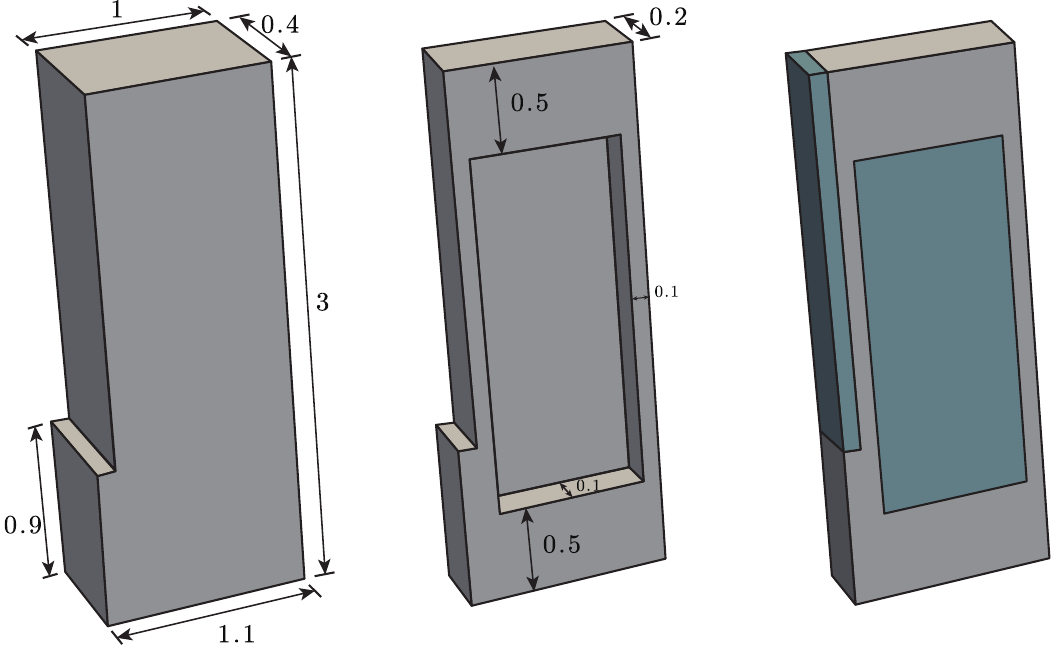}
    \caption{The geometry and size of a single cell of the pneumatic actuator, and the definition of a third medium.}
    \label{p11.s5.5.model1cell}
\end{figure}

For the last example, we consider the inflation simulation of a pneumatic actuator for a soft robot.
The geometry of the pneumatic actuator is shown in Fig.\ref{p11.s5.5.model} 
and the size for the single cell is given in Fig.\ref{p11.s5.5.model1cell}.
There are multiple air chambers in the actuator and the deformation is driven by pressurizing individual air chambers.
To effectively simulate the pneumatic actuation behavior as well as the possible self-contact, 
a third medium is introduced and filled within each air chamber and between each cell.
The size of the air chamber and the third medium are illustrated in Fig.\ref{p11.s5.5.model1cell}.
Due to symmetric, only 1/2 is considered.

\begin{figure}[htbp]
    \centering
    \includegraphics[width=1.0\textwidth]{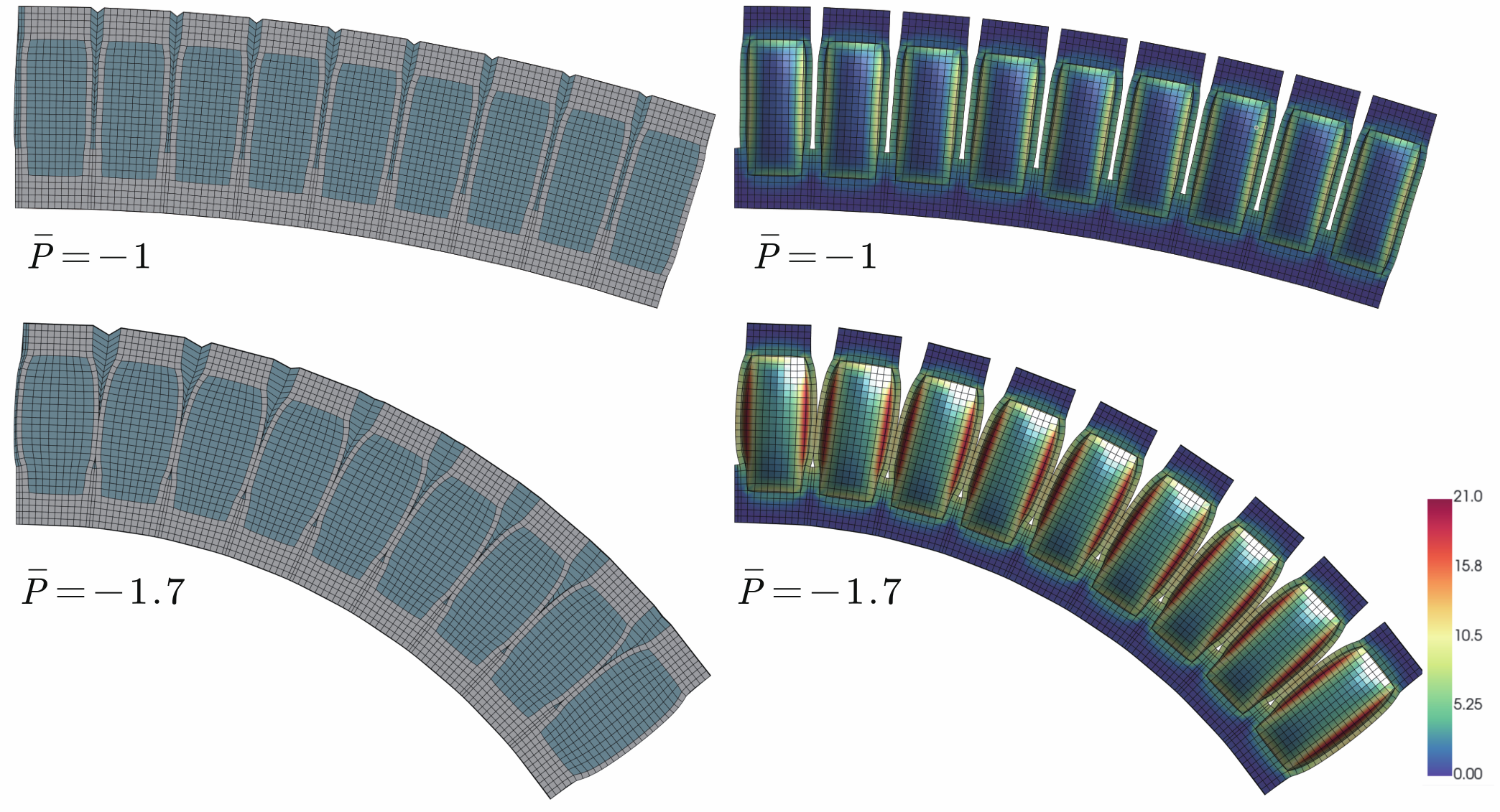}
    \caption{Von-Mises stress of the 1/2 pneumatic soft robot actuator under internal pressure.}
    \label{p11.s5.5.c}
\end{figure}

\begin{figure}[htbp]
    \centering
    \includegraphics[width=1.0\textwidth]{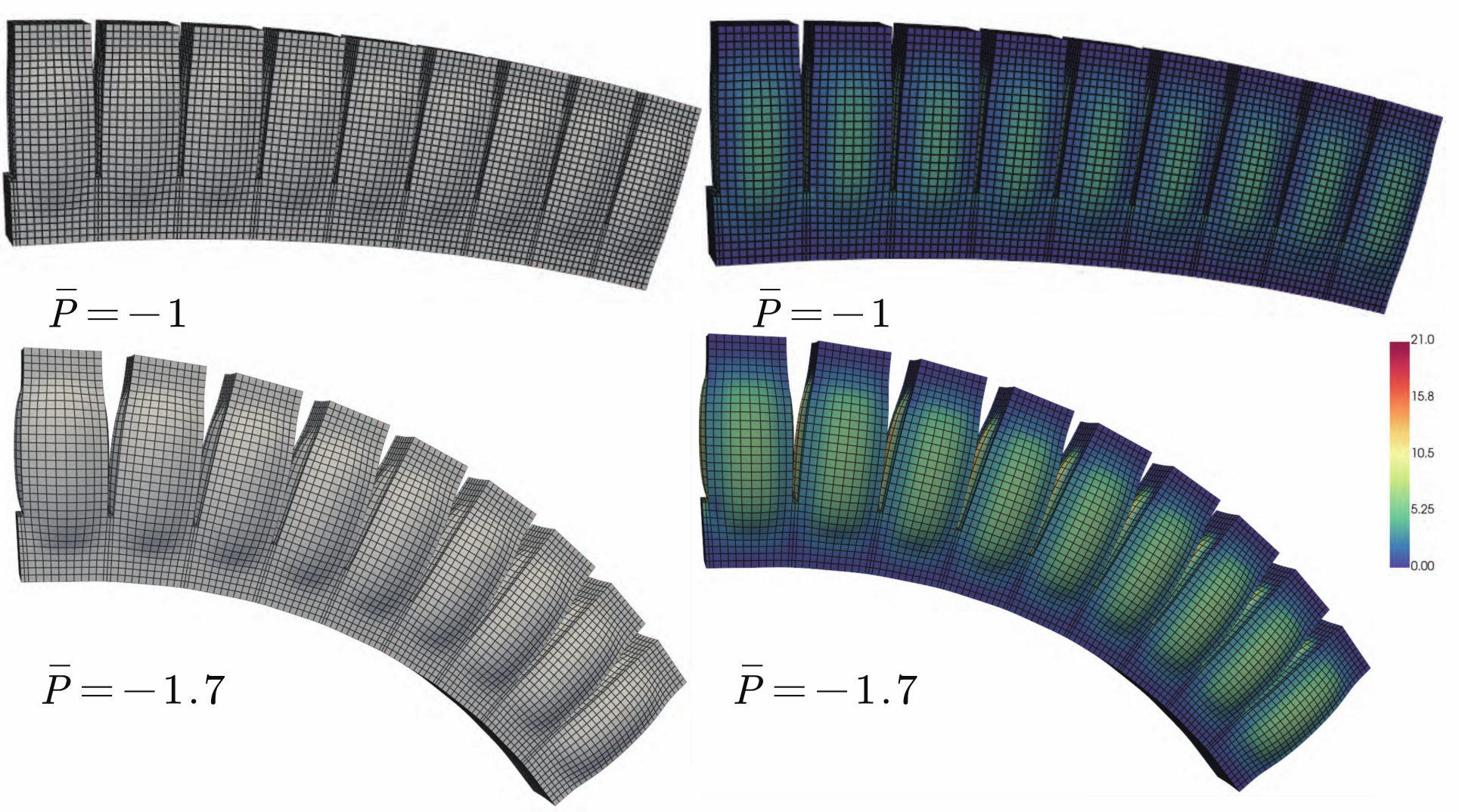}
    \caption{Deformation of the pneumatic soft robot actuator under internal pressure.}
    \label{p11.s5.5.def}
\end{figure}

The Neo-Hookean model as given in Eq.\eqref{p11.s2.strainEnergy} was used
with the shear modulus $K=20$ and the bulk modulus $\mu=10$.
The parameters for the third medium are selected as $\gamma = 1\times 10^{-5}$ and $\alpha_r = 100$.
For each chamber, an internal pressure $\bar{P} = -2$ is given.
The time step is selected as $\Delta t= 0.005$.
The von-Mises stresses of the 1/2 model for different internal pressures are depicted in Fig.\ref{p11.s5.5.c}.
It can be observed that, as the air pressure increases, 
each air chamber gradually expands and comes into contact, 
thereby leading to the deformation of the actuator. 
It is evident that the proposed third-medium framework can effectively handle this type of coupled pneumatic/contact problem.
The final deformation and stress of the entire model are shown in Fig.\ref{p11.s5.5.def}.
It should be noted that the current regularization term requires second-order elements, resulting in low computational efficiency. 
Future work could modify it to a first-order format, thereby significantly improving computational efficiency.

\section{Conclusions}
\label{p11.s6}
In this work, we developed a third medium contact framework for simulating frictionless hyperelastic contact and pneumatically actuated systems in three dimensions.
The third medium is modeled as a hyperelastic material with an additional pneumatic part in the strain energy density function.
A regularization term is introduced to improve the element quality of the third medium.
The finite element formulation is derived in detail, including the calculation of the tangent stiffness matrix.
Several numerical examples are presented to demonstrate the effectiveness of the proposed framework.
The results show that the third medium contact framework can effectively handle complex contact and self contact problems in 3D,
as well as simulate the deformation of pneumatically actuated systems.
The proposed framework can also be extended to other discretization methods, such as isogeometric analysis, virtual element method and meshless methods.
Future work may focus on some new first-order formulations and extending the framework to more complex geometries and loading conditions.

\section*{Acknowledgements}
\label{sec:acknowledgement}

The authors would like to acknowledge the support from the  Collaborative Research Project Grant (CRPG) by the Research Grants Council of Hong Kong (Project No. C7085-24GF).

\appendix
\section{Matrix forms of the constitutive tensors for the third medium}
\label{p11.appendix}
As discussed in section \ref{p11.s4.1},
the high-order tensors $\mathbb{A},\mathbb{B}$ and $\mathbb{C}$ in Eq.\eqref{p11.s4.1.w2} can be obtained by automatic differentiation
and then should be converted into matrix form $\hat{\mathbb{A}}, \hat{\mathbb{B}}$ and $\hat{\mathbb{C}}$, see Eq.\eqref{p11.s4.1.ddw}.
Here we give the components of the matrix forms of the high-order tensors:

\begin{equation}
	\hat{\mathbb{A}} = \begin{bmatrix}
		\mathbb{A}_{11111} & \mathbb{A}_{11121} & \mathbb{A}_{11131} & \mathbb{A}_{11112} & \cdots & \mathbb{A}_{11133}\\
		\mathbb{A}_{21111} & \mathbb{A}_{21121} & \mathbb{A}_{21131} & \mathbb{A}_{21112} & \cdots & \mathbb{A}_{21133}\\
		\mathbb{A}_{31111} & \mathbb{A}_{31121} & \mathbb{A}_{31131} & \mathbb{A}_{31112} & \cdots & \mathbb{A}_{31133}\\
		\mathbb{A}_{12111} & \mathbb{A}_{12121} & \mathbb{A}_{12131} & \mathbb{A}_{12112} & \cdots & \mathbb{A}_{12133}\\
		\mathbb{A}_{22111} & \mathbb{A}_{22121} & \mathbb{A}_{22131} & \mathbb{A}_{22112} & \cdots & \mathbb{A}_{22133}\\
		\mathbb{A}_{32111} & \mathbb{A}_{32121} & \mathbb{A}_{32131} & \mathbb{A}_{32112} & \cdots & \mathbb{A}_{32133}\\
		\vdots & \vdots & \vdots & \vdots & \ddots & \vdots\\
		\mathbb{A}_{33311} & \mathbb{A}_{33321} & \mathbb{A}_{33331} & \mathbb{A}_{33312} & \cdots & \mathbb{A}_{33333}\\
	\end{bmatrix},
\end{equation}
and 
\begin{equation}
	\hat{\mathbb{B}} = \begin{bmatrix}
		\mathbb{B}_{111111} & \mathbb{B}_{111211} & \mathbb{B}_{111311} & \mathbb{B}_{111121}& \mathbb{B}_{111221} & \mathbb{B}_{111321} & \cdots & \mathbb{B}_{111323} & \mathbb{B}_{111333}\\
		\mathbb{B}_{211111} & \mathbb{B}_{211211} & \mathbb{B}_{211311} & \mathbb{B}_{211121}& \mathbb{B}_{211221} & \mathbb{B}_{211321} & \cdots & \mathbb{B}_{211323} & \mathbb{B}_{211333}\\
		\mathbb{B}_{311111} & \mathbb{B}_{311211} & \mathbb{B}_{311311} & \mathbb{B}_{311121}& \mathbb{B}_{311221} & \mathbb{B}_{311321} & \cdots & \mathbb{B}_{311323} & \mathbb{B}_{311333}\\
		\mathbb{B}_{121111} & \mathbb{B}_{121211} & \mathbb{B}_{121311} & \mathbb{B}_{121121}& \mathbb{B}_{121221} & \mathbb{B}_{121321} & \cdots & \mathbb{B}_{121323} & \mathbb{B}_{121333}\\
		\mathbb{B}_{221111} & \mathbb{B}_{221211} & \mathbb{B}_{221311} & \mathbb{B}_{221121}& \mathbb{B}_{221221} & \mathbb{B}_{221321} & \cdots & \mathbb{B}_{221323} & \mathbb{B}_{221333}\\
		\mathbb{B}_{321111} & \mathbb{B}_{321211} & \mathbb{B}_{321311} & \mathbb{B}_{321121}& \mathbb{B}_{321221} & \mathbb{B}_{321321} & \cdots & \mathbb{B}_{321323} & \mathbb{B}_{321333}\\
		\vdots & \vdots & \vdots & \vdots& \vdots & \vdots & \ddots & \vdots & \vdots\\
		\mathbb{B}_{333111} & \mathbb{B}_{333211} & \mathbb{B}_{333311} & \mathbb{B}_{333121}& \mathbb{B}_{333221} & \mathbb{B}_{333321} & \cdots & \mathbb{B}_{333323} & \mathbb{B}_{333333}\\
	\end{bmatrix},
\end{equation}
\begin{equation}
	\hat{\mathbb{C}} = \begin{bmatrix}
		\mathbb{C}_{1111} & \mathbb{C}_{1121} & \mathbb{C}_{1131} & \mathbb{C}_{1112} & \cdots & \mathbb{C}_{1133}\\
		\mathbb{C}_{2111} & \mathbb{C}_{2121} & \mathbb{C}_{2131} & \mathbb{C}_{2112} & \cdots & \mathbb{C}_{2133}\\
        \mathbb{C}_{3111} & \mathbb{C}_{3121} & \mathbb{C}_{3131} & \mathbb{C}_{3112} & \cdots & \mathbb{C}_{3133}\\
		\mathbb{C}_{1211} & \mathbb{C}_{1221} & \mathbb{C}_{1231} & \mathbb{C}_{1212} & \cdots & \mathbb{C}_{1233}\\
        \mathbb{C}_{2211} & \mathbb{C}_{2221} & \mathbb{C}_{2231} & \mathbb{C}_{2212} & \cdots & \mathbb{C}_{2233}\\
        \vdots            & \vdots            & \vdots            & \vdots            & \ddots & \vdots\\
		\mathbb{C}_{3311} & \mathbb{C}_{3321} & \mathbb{C}_{3331} & \mathbb{C}_{3312} & \cdots & \mathbb{C}_{3333}\\
	\end{bmatrix}.
\end{equation}
\bibliographystyle{unsrt}
\bibliography{contact}



\end{document}